\documentclass[12pt,fleqn]{article}

\usepackage{amsmath}
\usepackage{amsfonts}
\usepackage{graphicx}
\usepackage{psfrag}

\renewcommand{\theequation}{\mbox{\arabic{section}.\arabic{equation}}}
\newcounter{saveeqn}[section]

\def\msection{\setcounter{equation}{0}%
\section}

\renewcommand{\thefigure}{\mbox{\arabic{figure}}} 
\newcounter{savefig} 
\newcommand{\alphfig}{ \setcounter{savefig}{\value{figure}}%
\addtocounter{savefig}{1} 
\setcounter{figure}{0}%
\renewcommand{\thefigure}{\mbox{\arabic{savefig}\alph{figure}}}} 
\newcommand{\resetfig}{\setcounter{figure}{\value{savefig}}%
\renewcommand{\thefigure}{\arabic{figure}}}

\newcommand{\ess}{\scriptscriptstyle}
\newcommand{\T}{\textstyle}

\newcommand{\nms}{\negmedspace}

\newcommand{\ko}{\kern0.035cm}
\newcommand{\co}{\kern0.02cm}
\newcommand{\br}{\kern0.04cm}
\newcommand{\bt}{\kern0.04cm}
\newcommand{\dpr}{\kern0.015cm}

\begin{document}

\renewcommand\figurename{\small Fig.$\nms$}

\title{\bf{Time evolution of the unstable two-fluid density fluctuations in Robertson-Walker Universes}}
\author{E. Gessner and H. Dehnen\\
Universit\"{a}t Konstanz\\
Fachbereich Physik\\
Fach M 568 \\
D-78457 Konstanz\\
e-mail: Heinz.Dehnen@uni-konstanz.de\\
Fax: 0049-7531-884864
} \maketitle

\newpage
\section*{Abstract}
We consider density fluctuations of a two-fluid model consisting
of hydrogen plasma and radiation prior to the cosmic hydrogen
recombination. As investigation method that of the dispersion
relations is applied, which have been derived from the general-relativistic sound-wave
equations taking into account
the coupling between plasma and radiation
carefully. We obtain growing unstable acoustic
modes within the mass range $2 \cdot 10^6 M_\odot < M < 6 \cdot 10^{12} M_\odot$.
In a second step the coupled differential equations for the
amplitudes of the unstable modes are integrated numerically with
respect to time where the integration extends from the initial time
prior to the hydrogen
recombination up to the present time. We find a significant increase of
the amplitudes up to 4 orders of magnitude, if the Universe is
described by a cosmological model with a positive cosmological
constant ($\Lambda \simeq 2,2 \cdot 10^{-56} \, \mbox{cm}^{-2}$) and
a positive curvature (Lema\^{\i}tre-Universe) without an essential amount
of cold dark matter. We conclude that the existence of galaxies
confirm these  statements.

\newpage
\msection{Introduction} In a previous paper (Ge{\ss}ner \& Dehnen
2000) we have investigated stability and instability of the
density fluctuations of a cosmic two-component fluid consisting of
hydrogen plasma and radiation for explaining the galaxy formation
using Newtonian cosmology and the method of the dispersion
relation. Although the investigation of cosmic density
fluctuations has been performed already by several authors (c.f.
the book of Coles \& Lucchin 1995), the essential new idea in our
paper was a careful treatment of the interaction of the plasma and
the radiation prior to the hydrogen recombination  taking into
account that the radiation pressure of the photons on the plasma
acts only on scales comparable or larger than the free path length
of the photons (c.f. Diaz-Rivera \& Dehnen 1998; Rose 1993, 1996).
In this way we were able to explain the mass-spectrum of the galaxies in
detail by the spectrum of the unstable fluctuation modes with a
lower and an upper mass limit of about $10^6 M_\odot$ and $10^{13}
M_\odot$ respectively.

An open question is however the time evolution of the unstable
fluctuations, especially its dependence on cosmological
parameters as the cosmological constant or quintessence,  the space-like
curvature and dark matter. Until now it is commonly assumed, that
most of the matter in the Universe exists in a nonbaryonic dark form,
in order to accelerate structure formation on large scales
(Peebles 1993, 2000; Hogan \& Dalcanton 2000; Hu et al. 2000). But
we will show, that such an assumption is not required.
Furthermore it is interesting to see, how the Newtonian
calculation of our previous paper follows from a
general-relativistic one.

Therefore in the present paper we formulate at first the
fundamental equations general-relativistic and
subsequently perform a first order perturbation calculation in order to
derive the sound-wave equations for the plasma- (matter-) and
radiation-densities. From these we proceed to the Newtonian limit
and demonstrate that this step is justified so long as the masses of the
matter fluctuations are smaller than $10^{19} M_\odot$ (see also
Bonnor 1957). For lower masses we determine using the method of the
dispersion relation once more the unstable fluctuation modes and
confirm the Newtonian results of our previous paper by the
general-relativistic approach.

In a second step we investigate the time evolution of the unstable
fluctuations by a numerical integration of the sound-wave
equations started from a point in time prior to
the recombination of the hydrogen
until the present time; the initial values for the numerical
calculations are chosen as identical with those of the unstable
modes according to the dispersion relation. We obtain the following
interesting result: The increase of the unstable fluctuation
amplitudes is, as it is well known (see eg. Weinberg 1972), not
described by an exponential function of time but rather by a power
law, since the coefficients of the sound-wave equations are
time-dependent due to the cosmic expansion, and it is argued, that
this increase is not sufficient for galaxy formation and that
therefore the assumption of dark matter is unavoidable. However
this situation changes drastically in favour of an exponential law
of time with an increase of the fluctuation amplitudes up to 4
orders of magnitude, if a positive cosmological constant or
quintessence is present and additionally
the space-like curvature is positive!
Therefore we conclude that the existence of
galaxies confirm these last two facts in agreement with the
results of W. Priester et al. (1995) and the new observation of an
acceleration of the cosmic expansion at the present time (Riess et
al. 1998).

Furthermore, the mode structure of our two-fluid model depends
essentially on the ratio of radiation and matter in the Universe.
If this ratio differs only slightly from the used value, the
modes and their time evolution change, which means that the
observed structures in the Universe reflect immediately the ratio
of radiation and matter. In this context it may be of interest,
that additional cold dark matter does not improve the comparison
with the observations in contradiction to the common opinion.

\msection{The fundamental general-relativistic equations for the
interacting two-fluid model}

Prior to the recombination of the hydrogen we consider the
electron-proton plasma and the photon gas as two components of the
cosmic fluid. The undisturbed homogeneous fluid is assumed to be
in thermodynamic equilibrium. The matter density of the plasma is
$\rho_{\rm p}$ and that of the radiation (photon gas) $\rho_{\rm r} = \sigma
\,T^4/c^2$ ($\sigma$  Stefan-Boltzmann constant, $T$  radiation
temperature). The plasma 4-velocity is denoted by $v_\mu \; (v^\mu\,
v_\mu = 1)$ and that of the radiation flow by $c_\mu \; (c^\mu \, c_\mu
= 1)$. Then the energy-momentum tensor for plasma and radiation is
given by:
\begin{equation}\label{2.1}
\hspace{-0.3cm}
\overset{{\rm\ess (p)}}{T}_{\nms \mu \nu} =
\left(\br \rho_{\rm p} c^2  + p_{\rm p} + p_{\gamma
\rightarrow {\rm e}} + p^R _{{\rm e} \leftarrow \gamma} \br\right)\,
v_\mu v_\nu -
\left(\br p_{\rm p} +
p_{\gamma \rightarrow {\rm e}} + p^R _{{\rm e} \leftarrow \gamma} \br\right)\,
g_{\mu \nu}
\end{equation}
and
\begin{equation}\label{2.2}
\hspace{-0.3cm}
\overset{{\rm\ess (r)}}{T}_{\nms \mu \nu} =
\left(\br \rho_{\rm r} c^2  + p_{\rm r} + p_{{\rm e} \rightarrow \gamma}
+ p^R_{\gamma \leftarrow {\rm e}}\br\right) \, c_\mu c_\nu -
\left(\br p_{\rm r} + p_{{\rm e} \rightarrow \gamma} +
p^R_{\gamma \leftarrow {\rm e}}\br\right) \, g_{\mu \nu}
\end{equation}
with the energy-momentum laws\footnote{$\;\;{\ess \|}\co\nu$  
means the
covariant partial derivative with respect to the coordinate
$x^\nu, $ and $ | \co \nu$ the usual one.}
\begin{equation}\label{2.3}
\hspace{-0.3cm}
\overset{{\ess\rm (p)}}{T}_{\nms\mu}{}^{\nu}{}_{{\ess \|}\co\nu}
= \overset{\ess ({\rm p}\leftarrow {\rm r})}{C_\mu} ,
\qquad \qquad
\overset{{\rm\ess (r)}}{T}_{\nms \mu}{}^{\nu}{}_{{\ess \|}\co\nu}
= \overset{\ess({\rm r} \leftarrow {\rm p})}{C_\mu} ,
\end{equation}
where
\begin{equation}\label{2.4}
\hspace{-0.3cm}
\overset{\ess ({\rm p} \leftarrow {\rm r})}{C_\mu} =
- \overset{\ess ({\rm r}\leftarrow {\rm p})}{C_\mu} =
\frac{4}{3}\,\rho_{\rm r} \,\frac{c^2}{l_\gamma}(\br c_\mu - v_\mu\br)
+ \rho_{\rm p} \, \frac{\overline{v}_{\rm e} c}{l_{\rm e}} \,
(\br c_\mu - v_\mu\br)
\end{equation}
describes the energy-momentum transfer between radiation and
plasma (kinetic energy head). Herein
\begin{equation}\label{2.5}
\hspace{-0.3cm}
l_\gamma = \frac{m}{\sigma_{{\ess\rm  Th}} \, \rho_{\rm p}} , \qquad\quad
l_{\rm e} = \frac{3}{4} \, \frac{m}{\sigma_{\ess\rm Th} \, \rho_{\rm r}}
\end{equation}
are the free path lengths of the photons $(l_\gamma)$ with respect
to the plasma and of the plasma particles $(l_{\rm e})$ with respect to
the photon gas. Insertion of (\ref{2.5}) into (\ref{2.4}) shows
that the second term of the right hand side of (\ref{2.4}) can be
neglected as long as the mean thermal velocity
$\overline{v}_{\rm e}$ of the electrons is significant smaller than the
velocity of light, i.e.
$\overline{v}_{\rm e} \ll c$. This is in fact
the case up to redshifts of the
order of $10^8$, that means for our purpose.
Of course, the interaction of the photons
with the plasma takes place essentially via the electrons and
therefore in (\ref{2.5}) $\sigma_{\ess\rm Th} = \frac{8 \pi}{3}
\bigl(\frac{e^2}{m_{\rm e} c^2}\bigr)^{\ess 2}$ is the Thomson scattering
cross-section of the electrons. On the other hand, the inertia of
the plasma is determined by the protons strongly connected with
the electrons, so that in (\ref{2.5}) $m$ denotes the proton mass.

Furthermore
\begin{equation}\label{2.6}
\hspace{-0.3cm}
p_{\rm r} = {\T\frac{1}{3}} \, \rho_{\rm r} \, c^2  , \qquad\qquad
p_{\rm p} = \rho_{\rm p} \, c_{\rm s}^{\,2}
\qquad \text{with} \quad
c_{\rm s}^{\,2} = 2 \; \T\frac{k_{\ess\rm B} \, T}{m}
\end{equation}
represent the pressures of radiation and matter in (\ref{2.1}) and
(\ref{2.2}) respectively and $c_{\rm s}$ is
the isothermal sound-velocity of the plasma\footnote{We prefer the
isothermal sound-velocity because the plasma is embedded in the
enormous heat-bath of the radiation. The ratio of the thermal
energy density of radiation and matter amounts to $3 \cdot
10^9$.}, which is considered as an ideal Boltzmann gas. The
remaining pressure-terms in (\ref{2.1}) and (\ref{2.2}) describe
the pressure interaction between plasma and radiation limited by
the free path lengths (\ref{2.5}) as follows:
\begin{equation}
\label{2.8}
\hspace{-0.6cm}
\text{(a)} \;\; p_{\gamma\rightarrow {\rm e}} = - p^R _{\gamma \leftarrow {\rm e}} =
e^{\frac{4}{9}\,l^{\,2}_\gamma \,\underline \Delta}\; p_{\rm r} ,
\qquad
\text{(b)} \;\;
p_{{\rm e} \rightarrow \gamma} = - p^R_{{\rm e} \leftarrow \gamma} =
e^{\frac{1}{3} \, l^{\,2}_{\rm e} \, \underline\Delta } \; p_{\rm p} ,
\end{equation}
where $\underline \Delta$ signifies the covariant 3-dimensional
Laplace-operator with respect to the rest frame of plasma and
radiation. Equation (\ref{2.8}a) represents the hydrostatic
pressure of the photon gas on the plasma (radiation pressure) and
its backreaction $(R)$ on the photons; relation (\ref{2.8}b)
describes analogously the hydrostatic pressure of the plasma on
the photon gas and its backreaction $(R)$ on the plasma. A
derivation of the expressions (\ref{2.8}) is given
in Appendix A. It is remarkable that the limitation factors in
(\ref{2.8}) are operator-valued, so that an
assumption of the special shape of the matter and radiation
fluctuations is not necessary in the beginning.

Finally, for the determination of the metric of space-time we use
Einstein's field equations of gravitation with the cosmological
constant $\Lambda$ ($G$ Newtonian gravitational constant):
\begin{equation}\label{2.10}
\hspace{-0.6cm}
R_{\mu \nu} = - \frac{8 \pi G}{c^4}\,\left(\br T_{\mu \nu} -
{\T\frac{1}{2}} \, T \, g_{\mu \nu}\br\right) + \Lambda \, g_{\mu \nu} ,
\end{equation}
where according to (\ref{2.1}), (\ref{2.2}) and (\ref{2.8}):
{\jot0.25cm
\begin{align}\label{2.11}
\hspace{-0.6cm}
T_{\mu \nu} = \,\overset{\rm\ess (p)} T_{\nms \mu \nu} +
\overset{\rm\ess (r)} T_{\nms\mu \nu}\,  =
&\; \rho_{\rm p} c^2 \, v_\mu v_\nu
- p_{\rm p} \, (\br g_{\mu \nu} - v_\mu v_\nu\br) +
\rho_{\rm r} c^2 \, c_\mu c_\nu -
\\
\hspace{-0.6cm}&
- p_{\rm r} \, (\br g_{\mu \nu } -c_\mu c_\nu\br) +
(\br p_{\gamma \rightarrow {\rm e}} -
p_{{\rm e} \rightarrow \gamma}\br)\, (\br v_\mu v_\nu - c_\mu c_\nu\br)
\nonumber
\end{align}}
\noindent\hspace{-0.3cm} with
\begin{equation}\label{2.12}
\hspace{0.5cm}
T_\mu {}^\nu {}_{{\ess ||}\co\nu} = 0
\end{equation}
guaranteed by (\ref{2.3}) and (\ref{2.4}). Due to the last term of the
right hand side of (\ref{2.11}) the reaction principle
in a relativistic theory is not fulfilled exactly.

\msection{Perturbation theory} For the investigation of the cosmic
matter and radiation fluctuations we start from the undisturbed
Robertson-Walker Universes with the line-elements in isotropic
space-like coordinates $x^i$:\footnote{Latin coordinate indices
run from 1 to 3.}
\begin{equation}\label{3.1}
\hspace{-0.5cm}
  d \overset{\ess 0}{s}{}^{\,2} = d{x ^4}^{\,2} -
  \frac{R^2 (t)}{\left(\br 1 + \frac{1}{4} \, \epsilon \, r^2 \br\right)^2}
\;  \delta_{ik} \; dx^i dx ^k , \;\quad
r^2 = \delta_{ik}\; x^i x ^k ,
  \;\quad x ^4 = ct,
\end{equation}
where $\epsilon = +1,0, - 1$ is valid for the case of positive,
  vanishing and negative curvature of the 3-dimensional space $t
  = $ const.. Subsequently we perform a first order perturbation
  ansatz as follows
{\jot-0.12cm
\begin{alignat}{5}
\hspace{-0.5cm}
g_{\mu \nu} &=& \, \overset{\ess 0}{g}_{\mu \nu} + \overset{\ess 1}{g}_{\mu \nu}(x^\alpha),
&\quad&
v_\mu &=& \, \overset{\ess 0}{v}_\mu + \overset{\ess 1}{v}_\mu (x^\alpha),
&\quad
c_\mu = \; \overset{\ess 0}{c}_\mu + \overset{\ess 1}{c}_\mu (x^\alpha), \nonumber\\
&&&&&&&\label{3.2}\\
\hspace{-0.5cm}
  \rho_{\rm p} &=\,& \overset{\ess 0}{\rho}_{\rm p} (t) \,+
  \overset{\ess 1}{\rho}_{\rm p} (x^\alpha),
  &\quad&
  \rho_{\rm r} &=& \,\overset{\ess 0}{\rho}_{\rm r} (t) \, +
  \overset{\ess 1}{\rho}_{\rm r} (x^\alpha)
  &\nonumber
\end{alignat}}
\noindent\hspace{-0.3cm} with
\begin{equation}\label{3.4}
\hspace{-0.5cm}
  \overset{\ess 0}{v}{}^\mu
  = \, \overset{\ess 0}{c}{}^\mu  =
  \, \overset{\ess 0}{v}_\mu = \, \overset{\ess 0}{c}_\mu = (0,0,0,1)
\end{equation}
and with respect to the normalization of $v_\mu$ and $c_\mu$:
\begin{equation}\label{3.5}
\hspace{-0.5cm}
\overset{\ess 1}{v}{}^4 = \, \overset{\ess 1}{c}{}^4 = - \overset{\ess 1}{v}_4 =
- \overset{\ess 1}{c}_4 = - \T\frac{1}{2} \, \overset{1}{g}_{44}.
\end{equation}
Herein we consider the first order deviations (denoted by 1) as
small compared with the background quantities (denoted by 0). The
undisturbed metric $\overset{\ess 0}{g}_{\mu\nu}$ is given by
(\ref{3.1}).

 Then starting from the field equations (\ref{2.10}), using
the energy-momen\-tum tensor (\ref{2.11}) and neglecting the
gas-pressure against the matter-density we obtain in the lowest order the
well known results (Friedman equations, $\,\dot {} =
\frac{\partial}{\partial x^4}$):
{\jot0.3cm
\begin{align}
\hspace{-0.5cm}
\alpha^2 &\equiv \frac{\dot R{}^2}{R^2} = \frac{8 \pi G}{3
c^2}\;
\bigl(\br\overset{\ess 0}{\rho}_{\rm p} +
\overset{\ess 0}{\rho}_{\rm r}\br\bigr) +
{\T\frac{1}{3}}\,
\Lambda - \frac{\epsilon }{R^2},
\label{3.6}\\
\hspace{-0.5cm}
\beta &\equiv  \frac{\ddot R}{R} = - \frac{4 \pi
G}{3 c^2}\;\bigl(\br\overset{\ess 0}{\rho}_{\rm p} +
2 \overset{\ess 0}{\rho}_{\rm r}\br\bigr) +
{\T\frac{1}{3}}\, \Lambda ,
\label{3.7}
\end{align}}
\noindent \hspace{-0.3cm}
and from the equations of motion (\ref{2.3}) (see also
(\ref{3.4})) it follows:
\begin{equation}\label{3.8}
\hspace{-0.5cm}
  \overset{\ess 0}{\rho}_{\rm p} \, R^3 = \mbox{const.}, \quad\qquad
  \overset{\ess 0}{\rho}_{\rm r} \, R^4 = \mbox{const.}.
\end{equation}

Concerning the density fluctuations of matter and radiation given
by $\overset{\ess 1}{\rho}_{\rm p}$ and
$\overset{\ess 1}{\rho}_{\rm r}$ we consider
only local ones in view of the fact, that the galaxy formation is
a local phenomenon in such a sense, that for the linear size of
the fluctuations, i.e. for their wave-length $\lambda$
the following relation holds:
\begin{equation}\label{3.9}
\hspace{-0.3cm}
\lambda = R \, r \ll R \quad \rightarrow \quad r \ll 1 .
\end{equation}
Thus for the computation of the cosmic density fluctuations we can
neglect all terms proportional to $\epsilon \, r, \epsilon\, r^2$ etc.
resulting from  (\ref{3.1}), so that the line-element can be
approximated by
\begin{equation}\label{3.10}
\hspace{-0.3cm}
  d s^2 = \underbrace{d {x^4}^{\,2} - R^2 (t) \, d \vec x ^{\,2}  }_{= \;
  \overset{\ess 0}{g}_{\mu \nu} \, dx^\mu d x^\nu} + \,
  \overset{\ess 1}{g}_{\mu \nu}(x^\alpha) \, d x^\mu d x^\nu ,
\end{equation}
including the first order perturbations of the metric, for which
we introduce the deDonder-gauge\footnote{$;\; \nu$ means the
covariant derivative with respect to the background metric
$\overset{\ess 0}{g}_{\mu \nu}$.}
\begin{equation}\label{3.11}
\hspace{-0.3cm}
   \psi_\mu{}^\nu{}_{\ko ;\co\nu} = 0 \quad \mbox{with} \;\,
  \psi_{\mu \nu} = \overset{\ess 1}{g}_{\mu \nu} - {\T\frac{1}{2}}
  \,\overset{\ess 1}{g}_{\alpha \beta} \, \overset{\ess 0}{g}{}^{\alpha  \beta}
  \, \overset{\ess 0}{g}_{\mu \nu}, \quad
  \psi_\mu {}^\nu = \psi_{\mu \alpha}\,  \overset{\ess 0}{g}{}^{\alpha \nu} .
  \end{equation}
After a (3+1)-decomposition the equations (\ref{3.11}) read for
the metric itself:
{\jot0.27cm
\begin{equation}
\label{3.12}
\begin{split}
&\hspace{-0.3cm}
R^{-2} \, \bigl(\br \overset{\ess 1}{g}_{mn\ko|\co n} - {\T\frac{1}{2}}\,
\overset{\ess 1}{g}_{nn\ko |\co m} \br\bigr) + {\T\frac{1}{2}}\,
\overset{\ess 1}{g}_{44\ko |\co m}
- \overset{\ess 1}{g}_{m4\ko |\co 4} -
3 \, \alpha \, \overset{\ess 1}{g}_{m4} = 0,
\\
&\hspace{-0.3cm}
R^{-2} \, \bigl(\br \overset{\ess 1}{g}_{4n\ko |\co n} - {\T\frac{1}{2}}\,
\overset{\ess 1}{g}_{nn \ko |\co 4} \br\bigr) - {\T\frac{1}{2}}\,
\overset{\ess 1}{g}_{44\ko |\co 4}
- 3 \, \alpha\, \overset{\ess 1}{g}_{44} = 0,
\end{split}
\end{equation}}
where the sum convention holds with respect to Latin indices.

Now we consider at first the density fluctuations following from
(\ref{2.3}) in first order perturbation calculation. Because the
fluctuation phenomenon of the plasma is non-relativistic
(sound-waves), it is appropriate to perform a (3+1)-decomposition.
Then we obtain from (\ref{2.3}) using the metric (\ref{3.10}) and
the decomposition (\ref{3.2}) the following equations of motion
and continuity \newline
\noindent for the plasma
{\jot0.27cm
\begin{equation}
\label{3.14}
\begin{split}
\hspace{-0.3cm}
\overset{\ess 0}{\rho}_{\rm p} \, \bigl(\br \overset{\ess 1}{v}{}^i{} _{|\co 4} +
\overset{\ess 1}{\Gamma}{}^{\, i} _{\, 44} +
2 \alpha \overset{\ess 1}{v}{}^i \br\bigr) =
- \T\frac{c^{\,2}_{\rm s}}{c^2} \, R^{-2} \,
\overset{\ess 1}{\rho}_{{\rm p}\ko |\co i}
&-
{\T\frac{1}{3}}\, R^{-2} \, e^{\frac{4}{9}l^{\,2}_\gamma \, \underline \Delta}
\; \overset{\ess 1}{\rho}_{{\rm r} \ko |\co i} + \\
&+ \T\frac{4}{3}\,
\frac{\overset{\ess 0}{\rho}_{\rm r}}{l_\gamma}\,
\bigl(\br \overset{\ess 1}{c}{}^i -
\overset{\ess 1}{v}{}^i \br\bigr),
\end{split}
\end{equation}}
\begin{equation}\label{3.15}
\hspace{-0.3cm}
\overset{\ess 0}{\rho}_{\rm p} \,
\bigl(\br \overset{\ess 1}{v}{}^i{}_{\ko |\co i} +
\overset{\ess 1}{\Gamma}{}^{\, i}_{\, 4i} \br \bigr) +
\overset{\ess 1}{\rho}_{{\rm p} \ko |\co 4}
+ 3 \, \alpha \, \bigl(\br \overset{\ess 1}{\rho}_{\rm p} + {\T\frac{1}{3}}\,
e^{\frac{4}{9}l^{\,2}_\gamma \, \underline\Delta} \;
\overset{\ess 1}{\rho}_{\rm r} \br \bigr) = 0
\end{equation}

\vspace{0.2cm}

\noindent and for the radiation
{\jot0.3cm
\begin{equation}
\label{3.16}
\begin{split}
\hspace{-0.3cm}
\T\frac{4}{3}\,\overset{\ess 0}{\rho}_{\rm r} \,
\bigl(\br \overset{\ess 1}{c}{}^i{}_{\ko |\co 4} +
\overset{\ess 1}{\Gamma}{}^{\, i}_{\, 44} +
\alpha \, \overset{\ess 1}{c}{}^i \br\bigr)
= &- {\T\frac{1}{3}}\, R^{-2}\, \bigl(\br 1 - e^{\frac{4}{9}l_\gamma^{\,2}
\,\underline\Delta}\br\bigr)\, \overset{\ess 1}{\rho}_{{\rm r}\ko |\co i} +\\
&+
\T\frac{4}{3}\,\frac{\overset{\ess 0}{\rho}_{\rm r}}{l_{\gamma}}
\,(\br\overset{\ess 1}{v}{}^i
 - \overset{\ess 1}{c}{}^i \br) - {\T\frac{4}{3}}\, \alpha
\, R^{-2}\, \overset{\ess 0}{\rho}_{\rm r} \, \overset{\ess 1}{g}_{4i} ,
\end{split}
\end{equation}}
\begin{equation}\label{3.17}
\hspace{-0.3cm}
  {\T\frac{4}{3}}\,\overset{\ess 0}{\rho}_{\rm r} \,
  \bigl(\br \overset{\ess 1}{c}{}^i{}_{\ko |\co i}
  +
  \overset{\ess 1}{\Gamma}{}^{\, i}_{\, 4 i} \br\bigr) +
  \overset{\ess 1}{\rho}_{{\rm r}\ko |\co 4} + 4 \,\alpha \,
  \bigl(\br 1 - {\T\frac{1}{4}}\,e^{\frac{4}{9}l_\gamma^{\,2}
   \,\underline\Delta}\br \bigr)\, \overset{\ess 1}{\rho}_{\rm r} = 0 ,
\end{equation}
where terms proportional to $c^{\,2}_{\rm s}/c^2 \leq
10^{-9}$ are neglected against comparable terms of the order of
One; the Christoffel symbols of the first order in (\ref{3.14})
to (\ref{3.17}) read:
\begin{equation}\label{3.18}
\hspace{-0.3cm}
  \overset{\ess 1}{\Gamma}{}^{\, i}_{\, 44} =
  {\T\frac{1}{2}}\, R^{-2}\,
  \bigl(\br \overset{\ess 1}{g}_{44\ko |\co i} -
  2 \overset{\ess 1}{g}_{4i\ko |\co 4}\br\bigr), \quad\quad
  \overset{\ess 1}{\Gamma}{}^{\, i}_{\, 4 i}  = R^{-2}\,
  \bigl(\br \alpha \overset{\ess 1}{g}_{ii} -
  {\T\frac{1}{2}}\, \overset{\ess 1}{g}_{ii\ko |\co 4}\br\bigr),
\end{equation}
whereas the operator $\underline \Delta$ has the form $\underline
\Delta = R^{-2}\, \Delta$, see Appendix A.

The sound-wave equations for the density perturbations are
obtained by the divergence of the equations of motion and the
subsequent substitution of the velocity derivatives by the
continuity equations. In this way we get with the use of the
Christoffel symbols (\ref{3.18}) and the deDonder-gauge (\ref{3.12})
from (\ref{3.14}) and (\ref{3.15})
{\jot0.32cm
\begin{align}
&\hspace{-0.9cm}
R^{-2}\,\bigl(\br \T\frac{c^{\,2}_{\rm s}}{c^2} \,
\delta_{{\rm p}\ko |\co i\co | \co i}
+ {\T\frac{1}{3}} \, d \,
e^{\frac{4}{9}l_\gamma^{\,2} \underline \Delta}\;
\delta_{{\rm r}\ko |\co i\co |\co i}\br\bigr)
- \delta_{{\rm p}\ko |\co 4\co |\co 4} -
\nonumber\\
&\hspace{-0.9cm}
-
2 \, \alpha \,
\Bigl[\br\delta_{{\rm p}\ko |\co 4} +
{\T\frac{1}{2}} \, d \, e^{\frac{4}{9}l_\gamma^{\,2} \underline
\Delta}\; \delta_{{\rm r}\ko |\co 4} +
{\T\frac{1}{2}} \, \overset{\ess 1}{\gamma}_{\co |\co 4}  + 2 \,
\overset{\ess 1}{g}_{44\ko |\co 4} \br\Bigr] -
\label{3.19}\\
&\hspace{-0.9cm}
- \T\frac{4}{3} \, \frac{d}{l_\gamma}\,
\Bigl[\br \delta_{{\rm p}\co |\co 4} - {\T\frac{3}{4}}\,
\delta_{{\rm r}\co |\co 4} +
{\T\frac{3}{4}} \alpha \, (\br 1 + {\T\frac{4}{3}}\,d \br)\,
e^{\frac{4}{9}l_\gamma^{\,2} \underline \Delta}\;\delta_{\rm r} \br\Bigr] - d
\,\bigl(\br\beta
+ {\T\frac{16}{9}}\,\alpha^2 \, l_\gamma^{\,2} \, \underline \Delta \br\bigr)
\, e^{\frac{4}{9}l_\gamma^{\,2} \underline \Delta} \; \delta_{\rm r} -
\nonumber\\
&\hspace{-0.9cm}
- \bigl(\br \alpha^2 + \beta\br\bigr) \,
\bigl(\br\overset{\ess 1}{\gamma} + 3 \,
\overset{\ess 1}{g}_{44}\br\bigr)  + {\T\frac{1}{2}}\, \bigl(\br R^{-2} \,
\,\overset{\ess 1}{g}_{44\co |\co i\co |\co i} -
\overset{\ess 1}{g}_{44\co |\co 4\co |\co 4}\br\bigr) = 0
\nonumber
\end{align}}
\noindent \hspace{-0.27cm} for the relative matter fluctuations
$\delta_{\rm p}$
 and from (\ref{3.16}) and (\ref{3.17})
{\jot0.32cm
\begin{align}
&\hspace{-0.9cm}
{\T\frac{1}{3}}\, R^{-2}\,
\bigl(\br 1 - e^{\frac{4}{9}l^{\,2}_\gamma\underline\Delta}\br\bigr) \,
\delta_{{\rm r}\ko |\co i\co |\co i} - \delta_{{\rm r}\ko |\co 4\co |\co 4} -
 \alpha \,\Bigl[\br \bigl(\br 1 -
e^{\frac{4}{9}l^{\,2}_\gamma\underline\Delta}\br\bigr) \,
\delta_{{\rm r}\ko |\co 4}
+ {\T\frac{4}{3}} \, \overset{\ess 1}{\gamma}_{\co |\co 4} +
{\T\frac{14}{3}} \, \overset{\ess 1}{g}_{44\ko |\co 4} \br\Bigr] +
\nonumber\\
&\hspace{-0.9cm}
+ l_\gamma^{\,-1} \, \Bigl[\br {\T\frac{4}{3}}\, \delta_{{\rm p}\ko |\co 4} -
\delta_{{\rm r}\ko |\co 4} + \alpha \,
\bigl(\br 1 + {\T\frac{4}{3}}\,d \br\bigr) \,
e^{\frac{4}{9}l^{\,2}_\gamma\underline\Delta}\; \delta_{\rm r} \br\Bigr]
+ \bigl(\br \beta + {\T\frac{16}{9}}\,\alpha^2 \,l^{\,2}_\gamma \,
\underline \Delta \br\bigr) \,
e^{\frac{4}{9}l^{\,2}_\gamma\underline\Delta} \; \delta_{\rm r} \ko -
\nonumber\\
&\hspace{-0.9cm}
- 4 \, \beta\,
\bigl(\br {\T\frac{1}{3}}\, \overset{\ess 1}{\gamma} + \overset{\ess 1}{g}_{44}
\br\bigr) +
{\T\frac{2}{3}}\,
\bigl(\br R^{-2} \, \overset{\ess 1}{g}_{44\ko |\co i\co |\co i} -
\overset{\ess 1}{g}_{44\ko |\co 4\co |\co 4} \br\bigr) = 0
\label{3.20}
\end{align}}
\noindent \hspace{-0.3cm} for the relative radiation fluctuations
$\delta_{\rm r}$ with the definitions
\begin{equation}\label{3.21}
\hspace{-0.3cm}
\delta_{\rm p} = \overset{\ess 1}{\rho}_{\rm p} \ko / \ko
                 \overset{\ess 0}{\rho}_{\rm p} , \qquad
\delta_{\rm r} = \overset{\ess 1}{\rho}_{\rm r} \ko / \ko
                 \overset{\ess 0}{\rho}_{\rm r} , \qquad
d = \overset{\ess 0}{\rho}_{\rm r}\ko /\ko\overset{\ess 0}{\rho}_{\rm p}, \qquad
\overset{\ess 1}{\gamma} = R^{-2}\,\overset{\ess 1}{g}_{ii}
\end{equation}
for the density contrasts, the ratio of the radiation and matter
background densities as well as the trace of the space-like metric
perturbations.

Finally we need the field-equations for the gravitational potentials
$\overset{\ess 1}{g}_{44}$ and $\overset{\ess 1}{\gamma}$.
 From (\ref{2.10})
and (\ref{2.11}) we obtain with the use of the decompositions
(\ref{3.2}) and the deDonder-gauge (\ref{3.12}) for $\mu = \nu =
4$:
{\jot0.3cm
\begin{align}
\hspace{-0.9cm}
R^{-2}\, \overset{\ess 1}{g}_{44\ko |\co i\co |\co i} -
\overset{\ess 1}{g}_{44\ko |\co 4\co |\co 4}
&-
2 \,\alpha \,\bigl(\br \overset{\ess 1}{\gamma}_{\co |\co 4} +
{\T\frac{5}{2}}\,\overset{\ess 1}{g}_{44\ko |\co 4}\br\bigr)
- 2 \, \alpha^2 \, \bigl(\br \overset{\ess 1}{\gamma} +
3 \, \overset{\ess 1}{g}_{44}\br\bigr)
- 2 \, \beta \, \overset{\ess 1}{\gamma} =
\nonumber\\
&
= \T\frac{8 \pi G}{c^2} \, \overset{\ess 0}{\rho}_{\rm p}\,
\left(\br\delta_{\rm p} + 2 \, d \, \delta_{\rm r}\br\right)
\label{3.22}
\end{align}}
\noindent \hspace{-0.3cm} and for the trace concerning $\mu, \nu = m, n$:
{\jot0.3cm
\begin{align}
  \hspace{-0.9cm}
  R^{-2}\,\overset{\ess 1}{\gamma}_{\co |\co i\co |\co i} -
  \overset{\ess 1}{\gamma}_{\co |\co 4\co |\co 4}
  &- 2 \ko \alpha \, \bigl(\br \overset{\ess 1}{g}_{44\ko |\co 4}  +
  {\T\frac{5}{2}}\, \overset{\ess 1}{\gamma}_{\co |\co 4}\br\bigr)
  + 6 \ko \alpha^2 \, \bigl(\br \overset{\ess 1}{\gamma}
  - \overset{\ess 1}{g}_{44}\br\bigr)
  - 6 \ko \beta \, \overset{\ess 1}{g}_{44} +
  4 \ko \T\frac{\epsilon}{R^2} \, \overset{\ess 1}{\gamma} =
\nonumber\\
  &= \T\frac{24 \pi G}{c^2} \, \overset{\ess 0}{\rho}_{\rm p} \,
  \bigl(\br \delta_{\rm p} + {\T\frac{2}{3}} \, d \,
  \delta_{\rm r} \br\bigr)
\label{3.23}
\end{align}}
\noindent \hspace{-0.3cm}
using also the zero order equations (\ref{3.6}) and (\ref{3.7}).

The 4 coupled wave-equations (\ref{3.19}), (\ref{3.20}),
(\ref{3.22}) and (\ref{3.23})
 are those of the self-gravitating two-fluid system in consequent first order perturbation. It
 is very interesting that using the deDonder-gauge the wave-operator applied to the metric perturbations
 $\overset{\ess 1}{g}_{44}$ and $\overset{\ess 1}{\gamma}$ does not only
 appear in the wave-equations for the metric (\ref{3.22}) and (\ref{3.23}) but also in the
 wave-equations of the matter and radiation fluctuations (\ref{3.19}) and (\ref{3.20}), so that
 it can be eliminated there. By doing this we obtain from
 (\ref{3.19}) and (\ref{3.20}) simultaneously substituting
 $\alpha^2$ and $\beta$ by (\ref{3.6}) and (\ref{3.7}):
{\jot0.35cm
\begin{align}
&\hspace{-0.95cm}
R^{-2}\ko\bigl(\br \T\frac{c_{\rm s}^{\,2}}{c^2} \,
\delta_{{\rm p}\co |\co i\co |\co i} + {\T\frac{1}{3}}\br
d\br e^{\frac{4}{9}l_\gamma^{\,2} \underline\Delta} \,
\delta_{{\rm r}\co |\co i\co |\co i} \co\bigr)
- \delta_{{\rm p}\co |\co 4\co |\co 4} -
 2 \co \alpha \ko
\Bigl[\ko \delta_{{\rm p}\co |\co 4} +
{\T\frac{1}{2}}\br d \br e^{\frac{4}{9}l_\gamma^{\,2}
\underline\Delta} \; \delta_{{\rm r}\co |\co 4} +
{\T\frac{3}{4}} \br \overset{\ess 1}{g}_{44\co |\co 4} \co\Bigr] -
\nonumber\\
&\hspace{-0.95cm}
- \T\frac{4}{3} \, \frac{d}{l_\gamma} \,\Bigl[\br \delta_{{\rm p}\ko |\co 4} -
{\T\frac{3}{4}}\, \delta_{{\rm r}\ko |\co 4} + {\T\frac{3}{4}} \,
\alpha \, (\br 1 + \frac{4}{3}\, d \br) \,
e^{\frac{4}{9}l^{\,2}_\gamma \underline\Delta}\;\delta_{\rm r} \br\Bigr] \br +
\nonumber\\
&\hspace{-0.95cm}
+ d \,
\Bigl[\br \T\frac{4\pi G}{3c^2}\, \overset{\ess 0}{\rho}_{\rm p} \,
\bigl(\br 1 + 2\,d - {\T\frac{32}{9}}\,(\br 1 + d\br) \, l^{\,2}_\gamma
\underline \Delta\br\bigr)
- {\T\frac{1}{3}} \,
\Lambda \, (\br 1 + {\T\frac{16}{9}} \, l^{\,2}_\gamma \, \underline \Delta \br)
\br +
\label{3.24}\\
&\hspace{6.05cm}
+ {\T\frac{16}{9}}\, \epsilon \,R^{-2} \, l^{\,2}_\gamma \, \underline \Delta
\br\Bigr]\, e^{\frac{4}{9} l_\gamma^{\,2}
\underline\Delta}\; \delta_{\rm r} \br +
\nonumber\\
&\hspace{-0.95cm}
+ \T\frac{4 \pi G}{c^2}\,\overset{\ess 0}{\rho}_{\rm p}\,
(\br \delta_{\rm p} + 2\, d\, \delta_{\rm r}\br)
 + \Bigl[\br \frac{4\pi G}{c^2} \, \overset{\ess 0}{\rho}_{\rm p}\,
 (\br 1+ 2\,d \br)-\Lambda \br\Bigr] \, \overset{\ess 1}{g}_{44} = 0
\nonumber
\end{align}}
\noindent \hspace{-0.3cm} and
{\jot0.35cm
\begin{align}
&\hspace{-0.95cm}
{\T\frac{1}{3}}\, R^{-2}\,
\bigl(\br 1 - e^{\frac{4}{9}l_\gamma^{\,2} \underline\Delta}\br\bigr) \,
\delta_{{\rm r}\ko |\co i\co |\co i} -
\delta_{{\rm r}\ko |\co 4\co |\co 4} - \alpha \, \Bigl[\br\bigl(\br 1-
e^{\frac{4}{9}l_\gamma^{\,2} \underline\Delta}\br\bigr) \,
\delta_{{\rm r}\ko |\co 4} + {\T\frac{4}{3}} \,
\overset{\ess 1}{g}_{44\ko |\co 4} \br\Bigr] \br +
\nonumber\\
&\hspace{-0.95cm}
+ l_\gamma^{\,-1} \, \Bigl[\br {\T\frac{4}{3}}\, \delta_{{\rm p}\ko |\co 4}
- \delta_{{\rm r}\ko |\co 4} +
\alpha \, (\br 1 + {\T\frac{4}{3}}\, d\br)\, e^{\frac{4}{9}l_\gamma^{\,2}
\underline\Delta} \; \delta_{\rm r} \br\Bigr] \br -
\nonumber\\
&\hspace{-0.95cm}
- \Bigl[\br \T\frac{4\pi G}{3c^2} \, \overset{\ess 0}{\rho}_{\rm p}\,
\bigl(\br 1 + 2 \, d  - {\T\frac{32}{9}}\,(\br 1 + d\br) \,
l^{\,2}_\gamma  \, \underline \Delta\br\bigr)
- {\T\frac{1}{3}} \,
\Lambda \, (\br 1 + {\T\frac{16}{9}} \, l_\gamma^{\,2} \, \underline \Delta \br)
\br +
\label{3.25}\\
&\hspace{5.85cm}
+ {\T\frac{16}{9}}\, \epsilon \, R^{-2} \, l^{\,2}_\gamma \, \underline
\Delta \br\Bigr] \, e^{\frac{4}{9}l_\gamma^{\,2} \underline\Delta} \,
\delta_{\rm r}
+
\nonumber\\
&\hspace{-0.95cm}
+ \T\frac{16 \pi G}{3 c^2} \, \overset{\ess 0}{\rho}_{\rm p} \,
(\br\delta_{\rm p} + 2 \, d\,  \delta_{\rm r}\br) + \frac{4}{3} \,
\Bigl[\br \frac{8 \pi G}{3c^2} \,
\overset{\ess 0}{\rho}_{\rm p}\, (\br 1 + d\br) - {\T\frac{1}{3}} \, \Lambda -
\frac{\epsilon}{R^2}\br \Bigr]  \, \overset{\ess 1}{\gamma}
\br +
\nonumber\\
&\hspace{-0.9cm}
+ 4 \, \Bigl[\br \T\frac{4 \pi G}{c^2} \, \overset{\ess 0}{\rho}_{\rm p} \,
(\br 1 + {\T\frac{4}{3}}\, d\br ) - \frac{\epsilon}{R^2} \br\Bigr] \,
\overset{\ess 1}{g}_{44} = 0.
\nonumber
\end{align}}
\noindent \hspace{-0.3cm} In earlier papers concerning galaxy formation on a
general-relativistic level
one has tried to solve the wave
equations for the metric perturbation and the sound-wave equations
for the density fluctuation simultaneously. However we will see
now, that this is not necessary for understanding the galaxy
formation; for this purpose the sound-wave equations (\ref{3.24}) and
(\ref{3.25}) for the matter and radiation fluctuations are totally
sufficient.

\msection{Newtonian limit}

The treatment of the set of the 4 first
order perturbation equations (\ref{3.22}) to  (\ref{3.25}) can
be simplified significantly by transition to the Newtonian limit
neglecting all terms proportional to $c^{-2}$ in (\ref{3.24}) and
(\ref{3.25}); with this approach the pairs of the equations (\ref{3.24}),
(\ref{3.25}) and (\ref{3.22}), (\ref{3.23}) can be solved
successively. From (\ref{3.22}) and (\ref{3.23}) one sees
immediately, that $\overset{\ess 1}{g}_{44}$ and
$\overset{\ess 1}{\gamma}$ are of the order of $c^{-2}$ and become
therefore important in (\ref{3.24}) and (\ref{3.25}) only for
fluctuations with masses $M \geq 10^{19} M_\odot$; a detailed
estimation of this value is given in Appendix B. Restricting
ourselves in view of the galaxy masses
to the only interesting lower mass range
 we can neglect in
(\ref{3.24}) and (\ref{3.25}) all terms proportional to
$\overset{\ess 1}{g}_{44}$, $\overset{\ess 1}{g}_{44\ko |\co 4}$ and
$\overset{\ess 1}{\gamma}$ (trans-Newtonian terms),\footnote{Note,
that the first order approximation is not identical with the
Newtonian limit.} so that the relevant sound-wave equations read
after additionally substituting $x^4 = ct$:
{\jot0.3cm
\begin{align}
&\hspace{-0.9cm}
R^{-2}\, \bigl(\br c^{\,2}_{\rm s} \, \delta_{{\rm p}\ko |\co i\co |\co i} +
{\T\frac{1}{3}}\, d \, c^2 \, e^{\frac{4}{9}l_\gamma^{\,2} \underline\Delta} \;
\delta_{{\rm r}\ko |\co i\co |\co i} \br \bigr)
- \delta_{{\rm p}\ko |\co t\co |\co t} -
2\,\alpha \,c \, \bigl(\br \delta_{{\rm p}\ko |\co t} +
{\T\frac{1}{2}}\, d \, e^{\frac{4}{9}l_\gamma^{\,2} \underline\Delta} \;
\delta_{{\rm r}\ko |\co t} \br\bigr) \br -
\nonumber\\
&\hspace{-0.9cm}
- \T\frac{4}{3}\, \frac{d\,c}{l_\gamma} \, \bigl[\br \delta_{{\rm p}\ko |\co t} -
{\T\frac{3}{4}}\, \delta_{{\rm r}\ko |\co t} +
{\T\frac{3}{4}}\, \alpha \, c \, (\br 1 + {\T\frac{4}{3}}\, d\br )\,
e^{\frac{4}{9}l_\gamma^{\,2} \underline\Delta}\,\delta_{\rm r}
\br\bigr] \br +
\nonumber\\
&\hspace{-0.9cm}
+ d \, \bigl[\br \T\frac{4 \pi G}{3} \, \overset{\ess 0}{\rho}_{\rm p} \,
\bigl(\br 1 + 2 \, d -
{\T\frac{32}{9}}\,(\br 1 + d\br) \, l^{\,2}_\gamma \,\underline \Delta\br\bigr)
- {\T\frac{1}{3}} \, \Lambda \, c^2 \,
(\br 1 + {\T\frac{16}{3}} \, l^{\,2}_\gamma \,
\underline \Delta\br) \br +
\label{4.1}\\
&\hspace{6.05cm}
+ \T\frac{16}{3} \, \epsilon \, \frac{c^2}{R^2 } \,
l^{\,2}_\gamma \, \underline \Delta \br\bigr] \, e^{\frac{4}{9}l_\gamma^{\,2}
\underline \Delta} \; \delta_{\rm r} \br  +
\nonumber\\
&\hspace{-0.9cm}
+ 4 \pi G \, \overset{\ess 0}{\rho}_{\rm p} \,
(\br \delta_{\rm p} + 2 \, d \, \delta_{\rm r}\br)
= 0
\nonumber
\end{align}}
and
{\jot0.3cm
\begin{align}
&\hspace{-0.9cm}
  \T\frac{1}{3} \, c^2 \, R^{-2} \,
  \bigl(\br 1 - e^{\frac{4}{9}l_\gamma^{\,2} \underline\Delta}\br\bigr) \,
  \delta_{{\rm r}\ko |\co i\co |\co i}
  - \delta_{{\rm r}\ko |\co t\co |\co t} - \alpha\, c \, \bigl(\br 1
- e^{\frac{4}{9}l_\gamma^{\,2} \underline\Delta} \br\bigr)  \,
\delta_{{\rm r}\ko |\co t}  \br +
\nonumber\\
&\hspace{-0.9cm}
+ \T\frac{c}{l_\gamma} \, \bigl[\br {\T\frac{4}{3}}\, \delta_{{\rm p}\ko |\co t} -
\delta_{{\rm r}\ko |\co t} + \alpha\, c\, (\br 1 + {\T\frac{4}{3}}\, d\br) \,
e^{\frac{4}{9}l_\gamma^{\,2} \underline\Delta} \; \delta_{\rm r} \br \bigr]
\br -
\nonumber\\
&\hspace{-0.9cm}
- \bigl[ \T\frac{4 \pi G}{3} \, \overset{\ess 0}{\rho}_{\rm p} \,
\bigl(\br 1 + 2\,d - {\T\frac{32}{9}}\, (\br 1 + d\br)\,
l^{\,2}_\gamma \, \underline \Delta \br\bigr) -
{\T\frac{1}{3}}\, \Lambda\, c^2\, (\br 1 + {\T\frac{16}{3}} \,
l^{\,2}_\gamma \, \underline \Delta\br)
\br +
\label{4.2}\\
&\hspace{5.75cm}
+ \T\frac{16}{9}\, \epsilon \,\frac{c^2}{R^2}\,l^{\,2}_\gamma \,\underline
\Delta\br\bigr] \, e^{\frac{4}{9}l_\gamma^{\,2} \underline\Delta}\;
\delta_{\rm r} \br +
\nonumber\\
&\hspace{-0.9cm}
+  \T\frac{16 \pi G}{3} \, \overset{\ess 0}{\rho}_{\rm p} \,
(\br\delta_{\rm p} + 2\, d \, \delta_{\rm r} \br) = 0.
\nonumber
\end{align}}
\noindent \hspace{-0.3cm}
It is very interesting, that these two equations for the matter
and radiation fluctuations  can be treated with the only use of
the background equations (\ref{3.6}) and (\ref{3.8}); the
equations (\ref{3.22}) and (\ref{3.23}) for $\overset{\ess 1}{g}_{44}$
and $\overset{\ess 1}{\gamma}$ are superfluous now! However these two
last wave-equations for the gravitational potentials
$\overset{\ess 1}{g}_{44}$ and $\overset{\ess 1}{\gamma}$ tell us that the
matter and radiation fluctuations described by (\ref{4.1}) and
(\ref{4.2}) are the source of gravitational waves immediately
correlated with the fluctuations (see also Appendix B and footnote 7).

The sound-wave equations (\ref{4.1}) and (\ref{4.2}) are identical
with those of the Newtonian cosmology (Ge{\ss}ner \&  Dehnen 2000),
apart from small modifications concerning the $\Lambda$- and
$\epsilon$-terms and a better consideration of the energy
conservation in (\ref{3.15}) and (\ref{3.17}). Thus it is
shown once more explicitly, that one reaches exactly the Newtonian
limit from general relativity by using the deDonder-gauge (Rose \&
Corona-Galindo 1991).

The integration of the equations (\ref{4.1}) and (\ref{4.2}) will
be performed in the following within two steps. First we
investigate prior to the recombination epoch the stable and
unstable oscillation modes with the method of the dispersion
relation, where the unstable modes yield the mass-spectrum for
the galaxies. Subsequently we use these results as initial values
for a numerical integration of the unstable modes up to the
present time in order to determine the amplitude increase
of the fluctuations during the evolution of the Universe. The last
step is required since  the
coefficients of the sound-wave equations (\ref{4.1}) and
(\ref{4.2}) are time-dependent, which implies
that the dispersion relation
method is only approximately valid for smaller time periods, see
(\ref{5.5}).

\msection{The dispersion relations}
\subsection{Theoretical approach}
As one can prove easily, prior to the recombination of the
hydrogen at the redshift $z_{\ess\rm R} = 1380 $ (see (\ref{5.18}) ff.) the
terms proportional to $\Lambda$ and $\epsilon$ in the equations
(\ref{3.6}), (\ref{3.7}) and (\ref{4.1}), (\ref{4.2}) can be
neglected in comparison with the $\overset{\ess 0}{\rho}_{\rm p}$-terms, since
at the recombination time the relations hold: $4\pi G\,
\overset{\ess 0}{\rho}_{\rm p}/\Lambda c^2 \simeq  1,5 \cdot 10^7$ and  $4
\pi G\, \overset{\ess 0}{\rho}_{\rm p}\, R^2 /c^2 \simeq 3 \cdot 10^2$ ($\Lambda
\approx 10^{-56} \, {\rm cm}^{-2}$). Then we obtain from (\ref{4.1}) and
(\ref{4.2}) with the eigenvibration ansatz
\begin{equation}\label{5.1}
 \delta_{\rm p} = \tilde A \, e^{i \, (\omega\, t - R \,\vec k \,\vec
 x)}, \quad\quad
 \delta_{\rm r} = \tilde B \, e^{i \, (\omega\, t - R \,\vec k \,\vec
 x)}
\end{equation}
with $(|\bt \vec k \bt| = k = 2 \pi /\lambda)$
\begin{equation}\label{5.2}
  k\,R = 2 \pi \bigl(\br\overset{\ess 0}{\rho}_{\rm p}
  \ko /\ko M\br\bigr)^{1/3} \, R
  = \mbox{const.}, \quad
  M = \overset{\ess 0}{\rho}_{\rm p} \, \lambda^3 = \mbox{const.}
\end{equation}
($\lambda$ wave-length and $M$ mass of the fluctuation) the
following algebraic equations for the complex-valued amplitudes
$\tilde A $ and $\tilde B$:
{\jot0.29cm
\begin{align}
\hspace{-0.95cm}  \tilde A  & \; \bigl[\br \omega^2  - i \, \omega\, c \,
\bigl(\br 2\, \alpha  +
  \T\frac{4}{3} \, \frac{d}{l_\gamma} \br\bigr)  -
  c_{\rm s}^{\,2} \, k^2 + 4 \pi G \,
  \overset{\ess 0}{\rho}_{\rm p} \br \bigr]  +
  \nonumber\\
 \hspace{-0.95cm} + \,
 \tilde B  & \; \bigl[\ko - i \ko \omega \ko c\, \bigl(\ko\alpha \ko d\ko
 e^{-\frac{4}{9}k^2 l^{\,2}_\gamma} - \T\frac{d}{l_\gamma}\ko\bigr) -
 {\T\frac{1}{3}}\ko d \ko c^2\ko k^2 \, e^{-\frac{4}{9}k^2 l^{\,2}_\gamma} +
 8 \pi G\ko d \ko\overset{\ess 0}{\rho}_{\rm p} -
   \label{5.3} \\
&\hspace{-0.45cm}
-  \T\frac{\alpha\ko c^2}{l_\gamma}\ko d \,
  (\co 1 + {\T\frac{4}{3}} \ko d \ko) \,
  e^{-\frac{4}{9}k^2 l^{\,2}_\gamma}
+ \frac{4\pi G }{3}\, \overset{\ess 0}{\rho}_{\rm p} \, d \,
 \bigl(\co 1 \!+\! 2\ko d \!+\!
 {\T\frac{32}{9}}\, (\co 1 \!+\! d\co)\, k^2\ko l^{\,2}_\gamma \ko\bigr)
 \,  e^{-\frac{4}{9}k^2 l^{\,2}_\gamma}\ko\bigr] = 0 \nonumber
\end{align}}
and
{\jot0.29cm
\begin{align}
\hspace{-0.95cm}  \tilde A  & \;
\bigl[\br \T\frac{4}{3}\, i \, \omega \, \frac{c}{l_\gamma }  +
\frac{16\pi G}{3}\,\overset{\ess 0}{\rho}_{\rm p} \br\bigr] +
\label{5.4}\\
\hspace{-0.95cm}+ \,
 \tilde B & \;
\bigl[\br \omega^2  -  i \ko \omega\ko c \, \bigl(\ko \alpha \, (\co 1  -
e^{-\frac{4}{9}k^2 l^{\,2}_\gamma} \ko)  +  \T\frac{1}{l_\gamma}\ko\bigr) -
  {\T\frac{1}{3}}\ko c^2\ko k^2 \,(\co 1  - e^{-\frac{4}{9}k^2
  l^{\,2}_\gamma}\ko) + \frac{32 \pi G}{3} \ko d \ko
  \overset{\ess 0}{\rho}_{\rm p} +
  \nonumber\\
&\hspace{-0.45cm}
  + \T\frac{\alpha \ko c^2}{l_\gamma}\,(\co 1 + {\T\frac{4}{3}}\ko d\br) \,
  e^{-\frac{4}{9}k^2 l^{\,2}_\gamma}
  - \frac{4\pi G}{3} \ko \overset{\ess 0}{\rho}_{\rm p} \,
  \bigl(\co 1 \!+\! 2\ko d \!+\!
  {\T\frac{32}{9}} \, (\co 1 \!+\! d\co)\, k^2\ko l^{\,2}_\gamma \ko\bigr) \,
  e^{-\frac{4}{9}k^2 l^{\,2}_\gamma} \ko\bigr] = 0 .
\nonumber
\end{align}}
\noindent \hspace{-0.29cm}
However, since the coefficients in (\ref{5.3}) and (\ref{5.4})
(see also (\ref{4.1}) and (\ref{4.2})) are time-dependent due to
the cosmic expansion, the frequencies $\omega$ become
time-dependent. Only if the coefficients in (\ref{5.3}) and
(\ref{5.4}) change slowly compared to the age of the Universe,
then also $\omega$ changes slowly with time as a parameter, and
the procedure with a nearly constant $\omega$ is a useful
approximation method for not too long times. In other words:
Deriving (\ref{5.3}) and (\ref{5.4}) the time derivative of
$\delta _{\rm p, \, r}\sim  e^{i \, \omega \, t}$ reads strictly $i(\omega +
\overset{\cdot}{\omega} \, t) e^{i \, \omega \, t}$, but the second term within
the bracket has been neglected under the assumption
\begin{equation}\label{5.5}
|\bt \overset{\cdot}{\omega} \bt | \, t \ll |\bt\omega\bt |.
\end{equation}
This condition for the applicability of the dispersion relation
method  will be proved retrospectively.

The two linear homogeneous equations (\ref{5.3}) and (\ref{5.4})
for the amplitudes $\tilde A $ and $\tilde B$ only have a
non-trivial solution, if the determinant of its coefficient matrix
vanishes. This condition results in a fourth-order equation for
the frequency $\omega(k)$ (dispersion relation)
\begin{equation}\label{5.6}
\omega ^4 + i \, \omega ^3 \, A - \omega ^2 \, B - i \, \omega \, C + D = 0
\end{equation}
with the (time-dependent) coefficients:
{\jot0.29cm
\begin{align}
\hspace{-0.8cm} A =& - 3 \,\alpha \, c \, \bigl(\br 1 - {\T\frac{1}{3}}\,
e^{-\frac{4}{9}k^2 l^{\,2}_\gamma}\br\bigr) -
\T\frac{c}{l_\gamma}\, \bigl(\br 1 + {\T\frac{4}{3}}\,d\br\bigr),
\label{5.7}\\
\hspace{-0.8cm}  B =&\; \T\frac{1}{3}\, c^2\, k^2 \,
\bigl(\br 1 - e^{-\frac{4}{9}k^2 l^{\,2}_\gamma} \br\bigr) +
2 \, \alpha c \, \bigl[\br \alpha c \,\bigl(\br 1  -
e^{-\frac{4}{9}k^2 l^{\,2}_\gamma} \br\bigr)
+ \frac{c}{l_\gamma} \, (\br 1 + {\T\frac{2}{3}}\, d\br)\br \bigr] -
 \nonumber\\
   &- 4 \pi G \, \overset{\ess 0}{\rho}_{\rm p} \,
   \bigl[\br 1 + {\T\frac{8}{3}}\, d
    - {\T\frac{1}{3}}\, \bigl(\br 1 + 2\,d + {\T\frac{32}{9}} \,(\co 1 + d\co)
    \,k^2\, l^2_ \gamma\br\bigr) \, e^{-\frac{4}{9}k^2 l^{\,2}_\gamma}
  \br\bigr]
  -  \nonumber\\
  &- \T\frac{\alpha \, c^2 }{l_\gamma}\,
  (\br 1 + {\T\frac{4}{3}}\,d\br)\, e^{-\frac{4}{9}k^2 l^{\,2} _\gamma},
\label{5.8}\\
\hspace{-0.8cm} C =& \; \alpha c \, \Bigl[ - \T\frac{2}{3} \, c^2 \, k^2 \,
\bigl(\br 1 - e^{-\frac{4}{9}k^2 l^{\,2}_\gamma}\br\bigr) +
4 \pi G \, \overset{\ess 0}{\rho}_{\rm p} \, \bigl[\br 1 +
{\T\frac{16}{3}}\, d -
(\br 1 + {\T\frac{4}{3}}\, d\br) \, e^{-\frac{4}{9}k^2 l^{\,2}_\gamma} -
\nonumber\\
&
- \T\frac{2}{3}\, \bigl(\br 1 + 2\,d + {\T\frac{32}{9}}\,
(\co 1 + d\co) \,k^2\, l_\gamma^{\,2}\br\bigr) \,
e^{-\frac{4}{9}k^2 l^{\,2}_\gamma}\br\bigr]
- 2 \frac{\alpha \,c^2}{l_\gamma}\,(\br 1 + {\T\frac{4}{3}}\,d\br) \,
e^{-\frac{4}{9}k^2 l^{\,2}_\gamma} \br\Bigr] +
\nonumber\\
&+
\T\frac{c}{l_\gamma}\,\bigl[\br 4 \pi G  \,
\overset{\ess 0}{\rho}_{\rm p} \, (\br 1 + 4\,d
+ {\T\frac{32}{9}}\, d^2 \br) - {\T\frac{4}{9}}\,d \,c^2\, k^2 \br\bigr],
\label{5.9}\\
\hspace{-0.8cm} D =&\; c^{\,2}_{\rm s} \,k^2
\bigl[\br \T\frac{1}{3}\,c^2 \, k^2 \,
\bigl(\br 1 - e^{-\frac{4}{9}k^2 l^{\,2}_\gamma}\br\bigr) -
\frac{\alpha \, c^2}{l_\gamma}\,(\br 1 + {\T\frac{4}{3}}\, d\br)
\, e^{-\frac{4}{9}k^2 l^{\,2}_\gamma} \br\bigr] -
\nonumber\\
&- \T\frac{4 \pi G }{3} \, \overset{\ess 0}{\rho}_{\rm p}\,
\bigl[\br c^2 \, k^2 \,
\bigl(\br 1 - (\br 1 + {\T\frac{4}{3}}\, d\br) \,e^{-\frac{4}{9}k^2
l^{\,2}_\gamma}\br\bigr) - 3 \, (\br 1 + {\T\frac{4}{3}}\, d\br)^2
\, \frac{\alpha \,c^2}{l_\gamma}\, e^{-\frac{4}{9}k^2 l^{\,2}_\gamma} +
\nonumber\\
&+ 4 \pi G \, \overset{\ess 0}{\rho}_{\rm p} \, (\br 1 + {\T\frac{4}{3}}\,d\br)
\bigl(\br
1 + 2\,d +
  {\T\frac{32}{9}}\, (\co 1 + d\co)\, k^2 \,l_\gamma ^2 \br\bigr)\,
  e^{-\frac{4}{9}k^2 l^{\,2}_\gamma} \br\bigr] .
\label{5.10}
\end{align}
In these expressions we have neglected again $c_{\rm s}^{\,2} k^2$-
against comparable $c^2 k^2$-terms. With the solutions $\omega(k)$ of
(\ref{5.6}) the amplitude ratio $r(k) = \tilde A /\tilde B$ is
determined by (\ref{5.3}) or (\ref{5.4}), which is in general
complex-valued including the phase relation between matter and radiation
fluctuations according to $r = |\bt r \bt| \, e^{i\phi}$.

Since the frequency $\omega$ as solution of (\ref{5.6}) is
complex-valued, we set:
\begin{equation}\label{5.11}
  \omega(k) = a(k) + i \, b (k), \qquad a(k), \,b(k) \in \mathbb{R} \, ;
\end{equation}
here $a(k)$ represents a real-valued
frequency connected with the period $T = 2 \pi /|\bt a\bt|$ and $\tau =
b^{-1}$ denotes a real-valued relaxation time of the fluctuations,
so that in view of (\ref{5.1}) $b < 0$ describes increasing and $b
> 0 $ decreasing amplitudes, i.e. $\tau < 0$  means growth times and $\tau > 0$ decay times.
Thus unstable modes are uniquely characterized by
\begin{equation}\label{5.12}
  a \equiv 0, \quad b < 0,
\end{equation}
which give rise to a kind of spinodal decomposition of the
homogeneous mixture of plasma and radiation.

Insertion of (\ref{5.11}) into (\ref{5.6}) and separation of real
and imaginary parts result in the two coupled relations for
$a(k)$ and $b(k)$:
{\jot0.3cm
\begin{align}
&\hspace{-0.5cm}
a^4 - a^2 \, (\br 6 \,b^2 + 3\,A\,b + B\br) + b^4 + A\,b^3
+ B\,b^2 + C\,b + D = 0 ,
\label{5.13}\\
&\hspace{-0.5cm}
a \, \bigl[\br a^2 (\br 4\, b + A\br) - 4\, b^3 - 3\, A\, b^2 -
2\, B\, b - C\br\bigr] = 0.
\label{5.14}
\end{align}}
\noindent \hspace{-0.3cm}
Because we will restrict ourselves in view of the galaxy formation
to the unstable modes defined by (\ref{5.12}) equation
(\ref{5.14}) is fulfilled identically, whereas equation
(\ref{5.13}) is reduced to:
\begin{equation}\label{5.15}
  b^4 + A\,b^3 + B\, b^2 +C\,b + D = 0,
\end{equation}
 the negative solutions of which yield the growth times as a function
 of $k$ or $M$  according to
 $k = 2 \pi \, \bigl(\br\overset{\ess 0}{\rho}_{\rm p} /M \br\bigr)^{1/3}$
 (see (\ref{5.2})). A detailed discussion of all
 solutions of (\ref{5.13}) and (\ref{5.14}) is given in our
 previous paper (Ge{\ss}ner \& Dehnen 2000). For the ratio
 $r = \tilde A /\tilde B$ we deduce
 from (\ref{5.4}) in the case of $a \equiv 0,$ see (\ref{5.12}):
{\jot0.29cm
\begin{align}
\hspace{-0.8cm} r = &
\Bigl[\br b^2 - b\,c \, \bigl(\br \alpha \, (\br 1 -
e^{-\frac{4}{9}k^2l_\gamma^{\,2}}\br ) + \T\frac{1}{l_\gamma }\br \bigr) +
{\T\frac{1}{3}}\,c^2\, k^2\, (\br 1 - e^{-\frac{4}{9}k^2l_\gamma ^{\,2}}\br) -
\frac{32 \pi G}{3} \, d \,  \overset{\ess 0}{\rho}_{\rm p} -
\nonumber\\
&-  \T\frac{\alpha\,
c^2}{l_\gamma}\,(\br 1 + {\T\frac{4}{3}}\, d\br)\,
e^{-\frac{4}{9}k^2l_\gamma ^{\,2}} +
\frac{4\pi G }{3} \, \overset{\ess 0}{\rho}_{\rm p}\, \bigl(\br 1 + 2\, d +
{\T\frac{32}{9}}\,(\co 1 + d\co)\, k^2\, l^{\,2} _\gamma \br\bigr)\,
e^{-\frac{4}{9}k^2l_\gamma ^{\,2}} \br \Bigr]
\nonumber\\
&/ \Bigl[\br \T\frac{16 \pi G}{3} \, \overset{\ess 0}{\rho}_{\rm p} -
\frac{4}{3} \, b \, \frac{c}{l_\gamma} \br\Bigr].
\label{5.16}
\end{align}}
\noindent \hspace{-0.3cm}
Evidently $r$ is real-valued, i.e. the
nonoscillating fluctuations of matter and radiation are in phase
($r > 0$, acoustic mode) or in phase opposition ($r < 0$, optical
mode).

For the numerical evaluation of (\ref{5.15}) and (\ref{5.16}) the
knowledge of the coefficients $A, B, C$ and $D$  prior to the
hydrogen recombination is required. To this end we reduce the
quantities in (\ref{5.7}) to (\ref{5.10}) with the help of the
"redshift"\footnote{Strictly speaking it holds $z $ = (redshift +
1).} $z = (\br R_{\rm\ess H} /R\br)$ to the today's background values
denoted by the index ${\rm H}$:
{\jot0.23cm
\begin{equation}
\label{5.17}
\begin{split}
&\hspace{-0.5cm}
\alpha \, c = \bigl[\br \T\frac{8\pi G}{3} \,
\overset{\ess 0}{\rho}_{\rm p {\ess H}} \, z^3
 \,(\br 1 + d_{\rm\ess H}\, z\br)\br\bigr]^{1/2} ,
\qquad k = k_{\rm\ess H} \, z ,
\qquad  l_\gamma = l_{\gamma {\rm\ess H}} \, z ^{-3} ,
\\
&\hspace{-0.5cm}
 c_{\rm s}^{\,2} = c_{\rm s {\ess H}}^{\,2}  \, z ,
 \qquad d = d_{\rm\ess H}  \, z,
 \qquad \overset{0}{\rho}_{\rm p} =
 \overset{\ess 0}{\rho}_{\rm p {\ess H}} \, z^3.
\end{split}
\end{equation}}
\noindent \hspace{-0.31cm}
In this context we take for the present background matter density
and the present background radiation temperature
the values
\begin{equation}\label{5.18}
\overset{\ess 0}{\rho}_{\rm p {\ess H}} = 2 \cdot 10^{-31} \, {\rm g/cm}^3 , \quad
\overset{\ess 0}{T}_{\rm\ess H} = 2,7 \, {\rm K},
\end{equation}
where the matter density corresponds to the baryonic one
determined by the abundance of the light chemical elements
generated in the cosmic nucleo\-synthesis. Accordingly, the hydrogen
recombination has taken place with regard to the Saha-equation at
the redshift $z = z_{\rm\ess R} = 1380$ and the today's background
quantities take the values (see (\ref{2.5}) and (\ref{2.6})):
{\jot0.145cm
\begin{equation}
 \label{5.19}
\begin{split}
    d_{\rm\ess H}  & \ko = \ko
   \overset{\ess 0}{\rho}_{\rm r {\ess H}} \co\co / \co\co
   \overset{\ess 0}{\rho}_{\rm p {\ess H}} =
    \sigma \overset{\ess 0}{T}{}^4 _{\rm\ess H} \co\co / \co\co
    (\br\overset{\ess 0}{\rho}_{\rm p {\ess H}} \, c^2 \br)
     = 2,24 \cdot 10^{-3},
     \\
    l_{\gamma {\rm\ess H}}  & \ko = \ko
       m \co\co / \co\co (\br \sigma_{\rm\ess Th} \,
       \overset{\ess 0}{\rho}_{\rm p {\ess H}} \br) =
     1,26 \cdot 10^{31} \, {\rm cm},
     \\
    c_{\rm s {\ess H}}^{\,2}  & \ko = \ko  2 \,
     k_{\rm\ess B} \overset{\ess 0}{T}_{\rm\ess H}
     \co\co / \co\co m
    = 4,46 \cdot 10^8 \, {\rm cm}^2 /{\rm sec}^2 ,
   \\
   k_{\rm\ess H}  & \ko = \ko  2 \pi
   \bigl(\br \overset{\ess 0}{\rho}_{\rm p {\ess H}} \co\co / \co\co
                           M \br\bigr)^{1/3} =
    2,92 \cdot 10^{-21} \, N^{-1/3} \, {\rm cm}^{-1} ,
\end{split}
\end{equation}}
\noindent \hspace{-0.3cm}
where $N$ is the number of solar masses ($M = NM_\odot$) of the
matter fluctuations.

With these data we have to solve the fourth-order equation
(\ref{5.15}) for $b$, which in general yields either 4 real-valued
solutions, or 2 real-valued solutions and 2 complex ones, or 4
complex-valued solutions; however, only the real-valued solutions
are relevant for our purpose. Thus we seek, with the help of a
standard computer algebra program, these solutions as functions of
$N$ for several values of $z$ or $s = z_{\rm\ess R}/z$ around and prior to
the recombination of the hydrogen $(z \approx z_{\rm\ess R} = 1380, s
\approx 1)$. Subsequently we calculate the real-valued amplitude
ratio (\ref{5.16}) also as function of $N$. The results are shown
in the Figs. 1 - 4. Finally we prove the condition (\ref{5.5}),
which reads for $b$  using (\ref{6.5}) and setting
$z = z_{\rm\ess R}/s$
\begin{equation}\label{5.20}
\hspace{-0.8cm}
  \left|\bt \frac{db/ds}{b} \br\right| \ll
  \T\frac{3}{2}s \br / \br \Bigl[\br (\br s + d_{\rm\ess H}\, z_{\rm\ess R} \br)
  (\br s - 2 \, d_{\rm\ess H}\, z_{\rm\ess R} \br) +
  2\, (\br d_{\rm\ess H} \, z_{\rm\ess R} \br)^{3/2} \,
  \sqrt{s + d_{\rm\ess H} \, z_{\rm\ess R}} \br\br\Bigr]
\end{equation}
for $s \leq 1$. The left hand side of
(\ref{5.20}) can be rewritten with the help of (\ref{5.15}) as
\begin{equation}\label{5.21}
\hspace{-0.7cm}
 \frac{db/ds}{b} =
\biggl[\ko\frac{dA}{ds}\br b^3 + \frac{dB}{ds}\br b^2 + \frac{dC}{ds}\br b +
          \frac{dD}{ds}\ko\biggr]\br
     \Bigl[\ko A \br b^3 + 2 \co B \br b^2 +
     3\co C\br b + 4 \co D\ko\Bigr]^{-1}
\end{equation}
and subsequently calculated for fixed $N$ at every point
(time/redshift) $s$ by the use of (\ref{5.7}) to (\ref{5.10})
and (\ref{5.17}).

\subsection{Results}
Fig. 1 shows qualitative courses of the solutions for $b$ along
with the ratio $r$ in the case of $a  \equiv 0$ (nonoscillating
modes) shortly before the hydrogen recombination. The mode $b_1 $
is a purely decaying optical one $(b > 0, r < 0)$ and has decay
times $\tau = b^{-1}$ of the order of $2,3 \cdot 10^3 \, {\rm y} > \tau >
1,8 \cdot 10^3 \, {\rm y}$, which is shown in Fig. 2, so that this mode is
damped out very rapidly at the recombination time of $4,8 \cdot
10^5 \, {\rm y}$ (see (\ref{6.5}) ff.). Of special interest is, however,
the mode $b_2$: first, a mode transition from an optical to an
acoustic mode lying in the damped range $(b
> 0)$ occurs at $2,01 \cdot 10^5 \, M_\odot$, so that at this point
the radiation fluctuation mode vanishes $(|\bt r\bt| \rightarrow
\infty$). Secondly this mode has an unstable acoustic region $(b <
0, r > 0)$ for $2,3 \cdot 10^6 \, M_\odot < M < 5,8 \cdot 10^{12}
\, M_\odot$, corresponding to scales between $219 \, {\rm ly}$ and $2,9 \cdot
10^4 \, {\rm ly}$ just before the recombination time. The ratio $r$ in this
unstable range is very large, i.e. of the order of $10^6$ up to
$10^8$ (see also Fig. 6), wherefore the radiation fluctuations are
very small compared to the matter fluctuations, so that it is
questionable, whether an observable influence on the background
radiation data exists.\footnote{The radiation density contrast is
correlated with a temperature fluctuation of the background
radiation according to $\delta_{\rm r} = 4 \, \Delta T/T$. A second
contribution to the temperature fluctuation results in consequence
of redshift from the variation of the  gravitational potential
according to $\Delta T/T = \overset{\ess 1}{g}_{44}/2$, see equ.
(\ref{B.3}).}

This instability region coincides exactly with that of the
observed masses of galaxies including globular clusters. Fig. 3
shows the behaviour of $b$ and $\tau = 1/b$ in detail for the mode
$b_2$ at $s = 1$; the growth time is of the order of $10^9 \, {\rm y}$, so
that the unstable growing mode needs an
additional strong increase during the later evolution of the
Universe for galaxy formation
(see also Bonnor 1956), which requires a positive
cosmological constant (quintessence) and positive curvature as
discussed in the next section. The whole instability valley as
function of "time" $s$ is illustrated in Fig. 4. The lower mass
limit is the Jeans mass, where gravity exceeds the plasma
pressure; the upper mass limit has its origin in the increasing
effective radiation pressure on the plasma with increasing size of
the fluctuations. The lower and the upper mass limit coincide at $
z \simeq 7600$, so that the instability valley ends with a mass
value of $2,3 \cdot 10^6 \, M_\odot$ at the time of $1,7 \cdot 10^4
\, {\rm y}$ $(s = 0,18; T = 20520 \, {\rm K})$. This fits approximately to that
time/redshift $(\br z \simeq [m^3 \, G/\pi \, k \,
T_{\rm\ess H} \, \rho_{\rm p {\ess H}}\br]^{1/4} \,
\sigma_{\rm\ess Th}^{-1/2})$, where the free path length of the photons
(upper mass limit) and the Jeans length (lower mass limit) become
identical, and which consequently represents the earliest time for
galaxy formation! All masses within the valley  show increasing
instability with time and also remain unstable after the hydrogen
recombination. Their further time evolution will be investigated
in the next section.

Finally Fig. 5 shows the applicability condition (\ref{5.20}) of
the dispersion relation  for the modes $b_1$ and $b_2$ and typical
masses versus $s$. In the case of the unstable growing mode $b_2$
the condition is fulfilled more or less, so that referring to this
our results presented above are decidedly useful.

\msection{Time evolution of the unstable fluctuation modes}
\subsection{Theoretical approach}
Since the dispersion relation method is valid only locally compared to the
timescale of the cosmic expansion, we investigate
the exact time evolution of the density contrasts with the ansatz
(instead of (\ref{5.1}))
\begin{equation}\label{6.1}
  \delta_{\rm p} = f(t) \, e^{-i\, R \, \vec k  \, \vec x } , \qquad
  \delta_{\rm r} = g (t)\, e^{-i \,  R \,\vec  k \,\vec x}
  \quad\qquad (R \, \vec k =
  \mbox{const.})
\end{equation}
and obtain from (\ref{4.1}) and (\ref{4.2}) the exact differential
equations for the amplitudes $f(t)$ and $g(t)$ ($\,\dot{} =
\frac{\partial}{\partial t}$):
{\jot0.25cm
\begin{align}
&\hspace{-0.8cm}
\ddot f + \bigl[\br 2\, \alpha \, c + \T\frac{4}{3} \,
d \, \frac{c}{l_\gamma}\br\bigr] \,\dot f +
d \, \bigl(\br \alpha \, e^{-\frac{4}{9}l^{\,2}_\gamma k^2}
- \frac{c}{l_\gamma}\br\bigr) \, \dot g +  c^{\,2}_{\rm s}  \, k^2 \, f +
\nonumber\\
&\hspace{-0.8cm}
+ d \, e^{-\frac{4}{9}l^{\,2}_\gamma k^2} \,
\Bigl[\br\T\frac{1}{3} \, c^2 \, k^2
+ \frac{\alpha \, c^2}{l_\gamma} \, (\br 1 + \frac{4}{3}\,d\br) -
\frac{4\pi G}{3}\,\overset{\ess 0}{\rho}_{\rm p} \, \bigl(\br 1 + 2 \, d +
\frac{32}{9}\,(\br 1 + d\br)\, k^2 \, l^{\,2} _\gamma\br\bigr) +
 \nonumber\\
&\hspace{2.65cm}
 + \T\frac{1}{3} \,\Lambda\, c^2\,
 \bigl(\br 1 - \frac{16}{9}\,k^2 \,l^{\,2} _\gamma\br\bigr) +
\frac{16}{9}\, \frac{\epsilon c^2}{R^2}\, k^2 \, l^{\,2} _ \gamma
\br\Bigr] \; g -
\nonumber\\
&\hspace{-0.8cm}
- 4 \pi G \, \overset{\ess 0}{\rho}_{\rm p} \, (\br f + 2 \, d \, g \br) = 0
\label{6.2}
\end{align}}
and
{\jot0.25cm
\begin{align}
&\hspace{-0.8cm}
\ddot g + \bigl[\br \alpha \, c\, \bigl(\br 1 -
e^{-\frac{4}{9}k^2 l^{\,2}_\gamma}\br\bigr) + \T\frac{c}{l_\gamma} \br\bigr] \,
\dot g -
\T\frac{4}{3} \, \frac{c}{l_\gamma}\, \dot f  +
\frac{1}{3} \, c^2 \, k^2 \, g -
\nonumber\\
&\hspace{-0.8cm}
- e^{-\frac{4}{9}k^2 l^{\,2}_\gamma} \,
\Bigl[\, \T\frac{1}{3}\,c^2 \,k^2  +
\frac{\alpha \, c^2 }{l_\gamma}\, (\br 1 + \frac{4}{3}\,d \br) -
\frac{4\pi G}{3}\,\overset{\ess 0}{\rho}_{\rm p} \,
\bigl(\br 1 + 2\,d +
\frac{32}{9}\,(\br 1 +
d\br)\, k^2\, l^{\,2}_ \gamma \br\bigr) +
 \nonumber\\
&\hspace{2.4cm}
+ \T\frac{1}{3}\,\Lambda \,c^2 \,\bigl(\br 1 -
\frac{16}{9}\,k^2 \,l^{\,2}_\gamma \br\bigr) +
\frac{16}{9}\, \frac{\epsilon c^2}{R^2}\,k^2\, l^{\,2} _ \gamma \br\Bigr] \; g -
\nonumber\\
&\hspace{-0.8cm}
-  \T\frac{16\pi G}{3}\, \overset{\ess 0}{\rho}_{\rm p} \,
  (\br f + 2 \,d \,g\br) = 0.
\label{6.3}
\end{align}}
\noindent \hspace{-0.35cm}
Additionally the differential equations (\ref{3.15}) and
(\ref{3.17}) can be integrated immediately with the use of
(\ref{6.1}); one gets for the flow-velocities in the Newtonian
limit
\begin{equation}\label{6.4a}
\hspace{-0.8cm}
  \overset{\ess 1}{v}{}^i =
  \bigl(\co \dot f  + \alpha \br d \br g \br
  e^{-\frac{4}{9}l_\gamma ^2 k^2} \co\bigr) \,
  \frac{k^i\br e^{-i\, R \,\vec k \,\vec x}}{i \,R\, k^2}
  , \quad\;
  \overset{\ess 1}{c}{}^i = {\T\frac{3}{4}} \br
  \bigl(\co \dot g - \alpha\br g \br e^{-\frac{4}{9}l_\gamma ^2
  k^2} \co\bigr) \,
  \frac{k^i \br e^{-i\, R \,\vec k \,\vec x}}{i\,R \,k^2},
\end{equation}
from which the real or imaginary part is to be taken
correspondingly to (\ref{6.1}).

Because the coefficients in (\ref{6.2}) and (\ref{6.3}) are known
as functions of $z$ in view of (\ref{5.17}) we substitute the time
$t$ through $z$ as an independent variable according to (c.f.
(\ref{3.6})):
\begin{equation}\label{6.4}
\hspace{-0.8cm}
\frac{dz}{z} = - \alpha \, c \, dt =
- \Bigl[\br \T\frac{8\pi G}{3}\,
\overset{\ess 0}{\rho}_{\rm p {\ess H}}\, z^3 \,
(\br 1 + d_H z \br) + \frac{1}{3} \, \Lambda \,
c^2 - \frac{\epsilon \, c^2 }{R^2 _H}\, z^2 \br\Bigr] ^{1/2} dt.
\end{equation}
Simultaneously we have taken into account the $\Lambda$- and
$\epsilon$-terms in (\ref{6.2}), (\ref{6.3}) and (\ref{6.4}) and
also in $\alpha$ given by (\ref{3.6}), in order to integrate
over a timescale starting prior to
the recombination epoch $(z > z_{\rm\ess R})$ up to the present time
$(z = 1)$. In this way the influence of the cosmological
constant $\Lambda$ and the curvature sign $\epsilon$ on the time
evolution of the amplitudes of the (unstable) fluctuations will be
investigated.

The differential equation (\ref{6.4}) for $z(t)$ cannot be
integrated exactly, so that the integration of (\ref{6.2}) and
(\ref{6.3}) with respect to $z$ is even necessary. Only for the
early stages of the Universe, where the $\Lambda$- and
$\epsilon$-terms on the right hand side of (\ref{6.4}) can be
neglected, i.e. for $z \geq z_{\rm\ess R}$, the integral can be given
explicitly as follows:
\begin{equation}\label{6.5}
\hspace{-0.8cm}
t = \T\frac{2}{3} \,
    \bigl(\br \frac{8\pi G}{3} \, \overset{\ess 0}{\rho}_{\rm p {\ess H}}
     \br\bigr)^{-1/2} \,
     z^{-3/2} \,
     \bigl[\, \sqrt{1 + d_{\rm\ess H} \, z}\,(\br1 - 2\,
d_{\rm\ess H} \, z\br) + 2\, (d_{\rm\ess H}\, z)^{3/2}\br\bigr].
\end{equation}
According to this solution (see also (\ref{5.18}), (\ref{5.19})) the
recombination of the hydrogen $(z_{\rm\ess R} = 1380)$ has happened at $t =
4,79 \cdot 10^5 \, {\rm y}$.

In the case, that the eqs. (\ref{6.2}) and (\ref{6.3})
are integrated over a timescale across
the hydrogen recombination, we have to consider
the decoupling of
 matter and radiation due to the hydrogen recombination.
 This means in detail that for $z < z_{\rm\ess R}$ holds
(instead of (\ref{5.17}))
\begin{equation}\label{6.6}
l_\gamma \rightarrow \infty \qquad (\br l_{\rm e} \rightarrow \infty\br), \qquad
c_{\rm s}^{\,2} = \frac{5}{3}  \, \frac{k_{\rm\ess B} \,
\overset{\ess 0}{T}_{\rm\ess H}}{m \, z_{\rm\ess R}} \, z^2 =
\T\frac{5}{6} \, c_{\rm s {\ess H}}^{\,2} \,
z_{\rm\ess R} ^{-1} \, z^2
\end{equation}
assuming an adiabatic cooling and an adiabatic sound-velocity for
the decoupled neutral hydrogen. In this context we have determined
the detailed behaviour of $l_\gamma$ (and $l_{\rm e}$) around
$z = z_{\rm\ess R}$
with the help of the actual ionisation degree according to the
Saha-equation.

For the numerical integration of (\ref{6.2}) and (\ref{6.3}) after
consideration of (\ref{6.4}) and (\ref{6.6}) we take as initial
values those of the (unstable) nonoscillating eigenmodes given by
the dispersion relation prior to the hydrogen recombination. For
this purpose we identify (\ref{5.1}) with (\ref{6.1}) setting:
\begin{equation}\label{6.7}
f = \tilde A \, e^{i\,\omega\, t} , \qquad g = \tilde B \, e^{i \,\omega\, t}
\end{equation}
and get with $\tilde A /\tilde B = r = |\bt r\bt|\, e^{i\,\phi}$
\begin{equation}\label{6.8}
  f = |\bt r \bt| \, e^{i\,(\br\omega \,t + \phi\br)} ,
  \qquad g = e^{i \,\omega \,t}
\end{equation}
up to an arbitrary common constant factor. The complex-valued
ansatz (\ref{6.7}) for solving the set of the linear homogeneous
differential equations (\ref{6.2}) and (\ref{6.3}) for the
real-valued density contrasts means, that the real and the
imaginary part of (\ref{6.8}) constitute the two independent real-valued
solutions of the system.

Considering at first the imaginary part we obtain after substitution
of $\omega$ according to (\ref{5.11})
{\jot0.25cm
\begin{align}
  f & =   |\bt r\bt| \, e^{-b\,t} \, \sin (\br a\,t + \phi\br), \qquad\quad
  g = e^{-b\,t} \, \sin a \,t,
  \nonumber\\
  \dot f & =  - |\bt r \bt| \, b \, e^{-b\,t} \, \sin (\br a\,t + \phi\br) +
  |\bt r\bt| \, a \, e^{-b\,t} \, \cos (\br a\,t + \phi\br),
  \label{6.9}\\
  \dot g & =  - b \, e^{-b\,t} \, \sin a \,t + a e^{-bt}\, \cos a\,t .
  \nonumber
\end{align}}
\noindent \hspace{-0.3cm}
Now for the nonoscillating modes with $a \equiv 0$
the amplitude ratio $r$ is real-valued
(see (\ref{5.16})), i.e. $\phi = 0, \pi.$ Herewith it
follows from (\ref{6.9})
\begin{equation}\label{6.10}
  f = 0, \qquad g = 0, \qquad \dot f = 0,
  \qquad \dot g = 0
\end{equation}
for every initial time $t$, so that the imaginary part leads only
to the trivial solution.

In the case of the real part it follows from (\ref{6.8})
{\jot0.25cm
\begin{align}
  f & = |\bt r \bt| \, e^{-b\,t} \, \cos (\br a\,t + \phi\br), \qquad
  g = e^{-b\,t} \, \cos a \,t,
   \nonumber\\
  \dot f & =  - |\bt r\bt| \, b \, e^{-b\,t} \, \cos (\br a\,t + \phi\br) -
  |\bt r\bt| \, a \, e^{-b\,t} \, \sin (\br a\,t + \phi\br),
  \label{6.11}\\
  \dot g & = - b \, e^{-b\,t} \, \cos a\, t - a \, e^{-b\,t} \, \sin a\,t .
\nonumber
\end{align}}
\noindent \hspace{-0.3cm} Restricting ourselves again
to the nonoscillating modes with $a \equiv 0$ we find for the
acoustic modes ($\phi = 0$)
\begin{equation}\label{6.12}
    f   =  |\bt r \bt|, \qquad g = 1,  \qquad
    \dot f =  - |\bt r\bt| \, b, \qquad
    \dot g = - b
\end{equation}
and for the optical modes $(\phi = \pi)$
\begin{equation}\label{6.13}
    f   = |\bt r \bt|, \qquad g = - 1,  \qquad
    \dot f  =  - |\bt r\bt| \, b, \qquad
    \dot g =  b
\end{equation}
as initial condition for every initial time $t$ unique up to a
common constant factor. The values of the 'frequency' $b$ and
the amplitude ratio $|\bt r \bt|$ for the
initial time $t$ connected with the initial "redshift" $z$
according to (\ref{6.5}) must be taken from the dispersion
relation (see (\ref{5.15}), (\ref{5.16}) and Fig. 1) after
choosing the mass $M$ of the fluctuation. Then the integration of
(\ref{6.2}) and (\ref{6.3}) can be performed numerically starting
before the hydrogen recombination at $z > z_{\rm\ess R} \; (= 1380)$, where for
technical reasons it is appropriate to use instead of $z$ the  new
independent variable $s = z_{\rm\ess R}/z$.

\subsection{Results and Conclusions}

The foregoing description of the numerical integration of
the differential equations
(\ref{6.2}) and (\ref{6.3}) for the fluctuation amplitudes
has been performed for several
cosmological models. In all cases we have chosen
$t = 1,25 \cdot 10^5 \, {\rm y} \; \hat = \; s = 0,5$
as starting time. The unstable modes are
all acoustic ones (see Fig. 1), i.e. the initial condition
(\ref{6.12}) holds. As masses we choose  $N = 4 \cdot 10^6$ and $N
= 7,9 \cdot 10^9$, which represent the boundary masses  of the
instability valley for the initial time; from the dispersion
relation we then get the values for $b$ and $|\bt r\bt|$, see Fig. 6.
Fig. 7 shows the behaviour of the matter fluctuation amplitudes
$f(s)$ up to the present time ($s = 1380)$ for several situations:

\vspace{0.2cm}

\noindent $\cdot$ case a) for $\Lambda = 2 \cdot
10^{-56} \, {\rm cm}^{-2}$, $\epsilon = 0$;

\vspace{0.05cm}

\noindent $\cdot$ case b) for $\Lambda = 0$,
$\epsilon = 0; $

\vspace{0.05cm}

\noindent $\cdot$ case c) for $\Lambda = 3,279 \cdot 10^{-56} \,
{\rm cm}^{-2}$, $\epsilon = +1$, $R_{\rm\ess H} =
3,388 \cdot 10^{28} \, {\rm cm}$, \\
\hspace*{1.7cm} $H = 93,4 {\rm km}/{\rm sec \, Mpc}; $

\vspace{0.05cm}

\noindent $\cdot$ case d) for $\Lambda = 2,206 \cdot 10^{-56}
\, {\rm cm}^{-2}$,
$\epsilon = +1$, $R_{\rm\ess H} = 3,306  \cdot 10^{28} \, {\rm cm}$, \\
\hspace*{1.6cm}  $H = 75 {\rm km}/{\rm sec \, Mpc};$

\vspace{0.2cm}

\noindent for the matter density holds
(\ref{5.18}) in all cases. Case c) is identical with a cosmological
model proposed by W. Priester
et al. (1995). Case d) represents the critical one, where
$\ddot R = 0, \dot R = 0$ at $s =
282$ (redshift 3,9). The last two cases c) and d) show clearly,
that a cosmological constant (or quintessence) together with
$\epsilon = +1$ promotes the galaxy formation very strongly by an
increase of the unstable matter density contrast up to 4 orders of
magnitude, whereas in the two first cases a) and b) the increase
is too small for galaxy formation. In this context it may be of
interest, that the strong increase of the matter density contrast
(region with $\alpha
  \simeq 0$, see below) occurs at redshifts of the
order of those of the quasars. On the other hand, a cosmological
constant or (nonbaryonic) cold dark matter alone, that means an
increase of $\alpha$
 (c.f. (\ref{3.6})),  reduces the tendency
for galaxy formation drastically, see case a). The physical reason
for this feature lies in the damping terms ($\alpha \,
\dot f$- and $\alpha \, \dot g$-terms) of the
differential equations (\ref{6.2}) and (\ref{6.3}). Only if these
terms go down for a longer time, and this happens according to
(\ref{3.6}) only by a compensation through the $\epsilon$-term
with $\epsilon = +1$, then the increase of the density contrast of
matter approaches to an exponential one with respect to time.
These results favour a cosmological model with a positive
cosmological constant and a positive curvature (Lema\^{\i}tre-Universe),
as also proposed by W. Priester et al. (1995)
-- however for other reasons --
and are in agreement with the analysis of the angular
power spectrum  of the background radiation (McGaugh 2000;
Overduin \& Priester 2001). Of course, this statement contradicts
the conjecture of a flat Universe concluded from the inflationary
cosmological model (c.f. the final remark). On the other hand, a
positive cosmological constant is also in accordance with the
observation of an acceleration of the cosmic expansion at the
present time (Riess et al. 1998).

Additionally, the behaviour of the density contrast $g(s)$ of the
radiation is given in Fig. 8 for the same cases as in Fig. 7;
after the decoupling from the matter due to the hydrogen
recombination, the radiation fluctuations decay finally very
rapidly as expected for the considered sizes. However, only in the
region of a very strong increase of the matter density contrast
also the radiation density contrast increases slightly in
consequence of gravitational coupling. But this property is no
longer that of an eigenmode; obviously the unstable
matter-radiation system leaves after some time the initial
eigenmode in consequence of the cosmic expansion.

In order to confirm the foregoing discussion of the influence of
nonbaryonic cold dark matter coupling only gravitationally the
calculations presented above are repeated, after in the
cosmological background equations (\ref{3.6}) and (\ref{3.7}) the
matter density $\overset{\ess 0}{\rho}_{\rm p}$ has been enlarged by
addition of several amounts of cold dark matter as follows:
$\overset{\ess 0}{\rho}_{\rm p} \rightarrow
\overset{\ess 0}{\rho}_{\rm p}\, (\br 1 + \delta \br)$ with
$\delta$ = 0,1; 0,25; 0,5; 1; 2,5; 5; 7,5;
10.\footnote{\samepage It is assumed, that the dark matter does
not participate in the fluctuations. Because of the gravitational
coupling this assumption is only approximately true.}
The results
for the matter and radiation fluctuation amplitudes are
illustrated in Fig. 9 and Fig. 10 for the critical case (with
$\dot R = \ddot R = 0 $), for which we
have obtained in Fig. 7 (case d) the strongest increase of the matter
fluctuation amplitudes; the initial time is again $s = 0,5$ and as
mass we choose $M = 1,8 \cdot 10^8 \, M_\odot$, which represents the
mean value of the unstable masses at the initial time. The
comparison with the case $\delta = 0$  shows, that even a small
amount of nonbaryonic cold dark matter reduces the increase of the
matter density fluctuations drastically, so that in view of the
existence of galaxies no essential amount of nonbaryonic cold dark
matter may be present in the Universe, which coincides also with
other new results, for instance with the analysis of the angular
power spectrum of the background radiation (for a review see
Overduin \& Priester 2001).

\msection{Final Remark}
 The results of our calculations compared with the observations of
 galaxies in the mass range of $2 \cdot 10^6 \, M_\odot < M <
 6 \cdot 10^{12} \,
 M_\odot$ favour a cosmological Robertson-Walker model with
a positive cosmological constant $(\Lambda \simeq 2,2 \cdot
 10^{-56}\, {\rm cm}^{-2})$ and a positive curvature (Lema\^{\i}tre-Universe),
 however without an essential
  amount of nonbaryonic cold dark matter. This is very satisfactory
 for two reasons: First of all, the age of the Universe will be of
 the order of approximately $30 \cdot 10^9 \, {\rm y}$ (c.f. Priester et
 al. 1995), so that no contradiction exists with respect to the age of
 the oldest stars. Secondly, a positive curvature connected with a
 finite (spherical) space is the only case, where the properties
 of the Universe as a whole, as e.g. homogeneity and isotropy can
 be proved empirically up to a certain extent. In all other cases
 this is totally impossible, wherefore the cosmological models
 with vanishing and negative (hyperbolic) curvature, i.e.
 infinite space, lie beyond the scope of natural sciences.
 Consequently, the interpretation of the results of the inflationary model as necessity for
 flatness of the Universe as a whole  is very problematic and not
 convincing (see also Ellis 1988; H\"{u}bner \& Ehlers 1991).

 \section*{References}
 \begin{itemize}
 \item Bonnor, W. B. 1956, Z. Astrophys. 39, 143
 \item Bonnor, W. B. 1957, Monthly Not. R.A.S. 117, 104
 \item Coles, P., \& Lucchin, F. 1995, Cosmology, The Origin and
 Evolution of Cosmic Structures, J. Wiley and Sons, Chichester
 \item Diaz-Rivera, L. M., \& Dehnen, H. 1998, Grav. and Cosmology
 4, 274
 \item Ellis, G.F.R. 1988, Class. Quant. Grav. 5, 891
\item Ge{\ss}ner, E. \& Dehnen, H. 2000, Grav. and Cosmology 6,
 194
 \item Hogan, C. \& Dalcanton, J. 2000, Phys. Rev. D62, 063511
 \item Hu, W. et al. 2000, Phys. Rev. Lett. 85, 1158
 \item H\"{u}bner, P. \& Ehlers, J. 1991, Class. Quant. Grav. 8, 333
 \item McGaugh, S. 2000, Int. J. Mod. Phys. A, Los Alamos preprint
 hep-ph/0010214
 \item Overduin, J. \& Priester, W. 2001, Naturwiss. 88, 229
 \item Peebles, P. 1993, Principles of physical cosmology,
 Princeton University Press
 \item Peebles, P. 2000, Astrophys. J. 534, L 127
  \item Priester, W. et al. 1995, Comments Astrophys. 17, 327, see
 also the literature cited there
\item Riess, A. 1998, Astron. J. 116, 1009
 \item Rose, B. \& Corona-Galindo M. 1991, GRG 23, 1317
 \item Rose, B. 1993, GRG 25, 503
 \item Rose, B. 1996 Il Nuovo Cimento 111, 1125
 \item Weinberg, S. 1972, Gravitation and Cosmology, John Wiley and
 Sons, New York
\end{itemize}

 \section*{Appendix A \newline
 Derivation of the radiation pressure in an optically thin
 medium}
\setcounter{section}{1}
 \setcounter{equation}{0}
\renewcommand{\theequation}{\Alph{section}.\arabic{equation}}

 We consider photons confined by a plasma. The usual radiation
 pressure $p_{\rm r}$ on the plasma is then proportional to the
 radiation energy density $u = \rho_{\rm r}\, c^2$:
\begin{equation}\label{A.1}
  p_{\rm r} = \T\frac{1}{3} \, u.
\end{equation}
However, this is only correct if the plasma is optically thick, so
that $u$ remains nearly constant with time. This does not apply in
an optically thin plasma because of the stronger diffusion of the
photons and their energy.

Thus we have to derive in a first step the relativistic diffusion
equation for the radiation energy. For this we start from the
energy-momentum tensor for radiation $(c_\mu \, c^\mu = 1)$
\begin{equation}\label{A.2}
  T_\mu {}^\nu = u \, c_\mu \, c^\nu -
       p_{\rm r}\, (\br \delta _\mu{} ^\nu - c_\mu \,
  c^\nu \br)
\end{equation}
and obtain with (\ref{A.1}) and $T_\mu {}^\nu {}_{\ess ||\nu} = 0$ and
subsequent multiplication with $c^\mu$:
\begin{equation}\label{A.3}
(\br u \, c_\mu \br)_{\ess ||\nu} \, g^{\mu \nu} =
\T\frac{1}{4} \, u_{|\co \mu} \, c^\mu .
\end{equation}
With regard to the decompositions (\ref{3.2}) up to (\ref{3.5})
equation (\ref{A.3}) is fulfilled in the lowest order; but in the
first order we obtain after a (3+1)-decomposition
\begin{equation}\label{A.4}
\T\frac{4}{3} \, (\br\overset{\ess 0}{u}\, \overset{\ess 1}{c}_m
\br)_{|\co n} \, \overset{\ess 0}{g}{}^{mn} +
\overset{\ess 1}{u}_{|\co 4} + 4 \, \alpha \,
\overset{\ess 1}{u} = 0,
\end{equation}
where all terms proportional to $c^{-2}$ are neglected
simultaneously (Newtonian limit). With the assumption of Fick's
law for the diffusion flow
\begin{equation}\label{A.5}
\overset{\ess 0}{u} \, \overset{\ess 1}{c}_m =
\T\frac{1}{3} \, l_\gamma \,
\overset{\ess 1}{u}_{|\co m}
\end{equation}
($\lambda_\gamma$ free path length of the photons) and assuming a
rapid diffusion compared with the cosmic expansion we get from
(\ref{A.4}) the following diffusion equation for the radiation
energy:
\begin{equation}\label{A.6}
  R^{-2} \, \overset{\ess 1}{u}_{|\co m\co |\co m} -
  \T\frac{9}{4 \, l_\gamma \, c} \,
  \overset{\ess 1}{u}_{|\co t} = 0.
\end{equation}
Now, the effective radiation pressure
$\overset{\ess 1}{p}_{\gamma \rightarrow {\rm e}}$ on the plasma
is determined by the photon energy
density remaining after a mean collision time $\tau = l_\gamma/c$
of the photons with the plasma electrons. Thus we obtain
\begin{equation}\label{A.7}
\overset{\ess 1}{p}_{\gamma \rightarrow {\rm e}}(t) =
\T\frac{1}{3} \, \overset{\ess 1}{u}(t +\tau), \qquad \tau \ll t
\end{equation}
with the Taylor expansion
\begin{equation}\label{A.8}
  \overset{\ess 1}{p}_{\gamma \rightarrow {\rm e}}(t) =
  \T\frac{1}{3} \, \sum^\infty _{n = 0}  \, \frac{1}{n!} \,
  \overset{\ess 1}{u}{}^{(n)}(t) \, \tau ^n.
\end{equation}
On the other hand, from the diffusion equation (\ref{A.6}) it
follows by iteration and under the assumption, that $R(t) \simeq
R(t + \tau)$:
\begin{equation}\label{A.9}
\overset{\ess 1}{u}{}^{(n)}(t) =
\bigl(\br \T\frac{4}{9} \, c \, l_\gamma \, R^{-2} \Delta \br\bigr)^n
\, \overset{\ess 1}{u}(t).
\end{equation}
Insertion of (\ref{A.9}) into (\ref{A.8}) gives with $\tau =
l_\gamma/c$:
{\jot0.3cm
\begin{equation}\label{A.10}
  \begin{split}
    \overset{\ess 1}{p}_{\gamma \rightarrow {\rm e}} (t) &=
    \T\frac{1}{3} \,
    \sum^\infty _{n = 0} \, \frac{1}{n!} \,
    \bigl(\br \T\frac{4}{9}\, \frac{l_\gamma^{\,2}}{R^2 }\,
    \Delta\br\bigr)^n  \, \overset{\ess 1}{u}(t)
     = \\
    &=
    \T\frac{1}{3} \, e^{\frac{4}{9}l_\gamma^{\,2} \underline \Delta}
    \; \overset{\ess 1}{u}(t) =
e^{\frac{4}{9}l_\gamma^{\,2} \underline \Delta} \;
\overset{\ess 1}{p}_{\rm r} ,
\qquad\qquad \underline \Delta = R^{-2} \, \Delta.
  \end{split}
\end{equation}}
\noindent \hspace{-0.3cm}
The same consideration is valid for usual nonrelativistic gas
particles, where the diffusion equation for the particle density
is identical with (\ref{A.6}) after substituting the factor $9/4$
by 3. The operator $e^{\frac{4}{9}l_\gamma^{\,2} \underline \Delta}$
in (\ref{A.10}) takes into account that the pressure is determined
by the particles (photons) remaining after a mean collision time
and that it takes effect only on scales larger than the free path
length $l$ of the particles under consideration.

\section*{Appendix B \newline Estimation of the order of magnitude of the trans-Newtonian
$\overset{\ess 1}{g}_{44}$- and $\overset{\ess 1}{\gamma}$-terms}

\setcounter{section}{2}
\setcounter{equation}{0}
\renewcommand{\theequation}{\Alph{section}.\arabic{equation}}

The differential equation (\ref{3.22}) for the gravitational
potential $\overset{\ess 1}{g}_{44}$ reads in its leading terms with
respect to $c^{-2}$, i.e. in the near zone (Newtonian limit):
\begin{equation}\label{B.1}
  R^{-2} \, \overset{\ess 1}{g}_{44\co|\co i\co |\co i} =
  \T\frac{8\pi G}{c^2} \, \overset{\ess 0}{\rho}_{\rm p}\,
  (\br \delta_{\rm p} + 2 \, d  \, \delta_{\rm r}\br).
\end{equation}
A similar Poisson-equation is valid for $\overset{\ess 1}{\gamma}$,
wherefore we restrict ourselves to the discussion of the
$\overset{\ess 1}{g}_{44}$-terms. With the ansatz analogous to
(\ref{5.1}), (\ref{6.1}):
\begin{equation}\label{B.2}
\overset{\ess 1}{g}_{44} = \tilde C \, e^{i\,  (\omega \,t - R \,\vec k\,\vec x)}
, \qquad \overset{\ess 1}{g}_{44} = h(t) \, e^{-i\,R \vec k \,\vec x}
\end{equation}
we obtain from (\ref{B.1}) immediately the gravitational
potential:
\begin{equation}\label{B.3}
  \overset{\ess 1}{g}_{44} = - \T\frac{8\pi G}{c^2 k^2 }\,
  \overset{\ess 0}{\rho}_{\rm p}\,
  (\br \delta_{\rm p} + 2 \,d \,\delta_{\rm r}\br).
\end{equation}
A comparison with the equations (\ref{3.24}), (\ref{3.25}) shows
now, that the $\overset{\ess 1}{g}_{44}$-terms can be neglected so
long as
\begin{equation}\label{B.4}
  \T\frac{8\pi G}{c^2 k^2} \; \overset{\ess 0}{\rho}_{\rm p} =
  \frac{2G}{\pi c^2} \;
  \overset{\ess 0}{\rho}_{\rm p{\ess H}}{}^{1/3} \, M_\odot^{2/3}
  \, N^{2/3} \, z \; \ll \; 1
\end{equation}
is valid. For the hydrogen recombination epoch ($z_{\rm\ess R} = 1380$) one
finds with the data (\ref{5.18}) $N \ll 6,8 \cdot 10^{19}$ for
neglecting the trans-Newtonian terms in (\ref{3.24}),
(\ref{3.25}).  For decreasing $z$-values the critical $N$-value
even increases.
 \newpage
\section*{Figure Captions}

{\bf{Fig. 1}}: Qualitative courses of the frequency
$b$ (a) and the amplitude ratio $r$ (b) for the
nonoscillating modes $(a \equiv 0)$ versus $\lg N$. There exist
two main modes $b_1$ and $b_2$, where the sign characterizes the
optical ($-$) and the acoustic (+) ones. The numbers refer to the
recombination epoch $(s = 1)$.

\noindent {\bf{Fig. 2}}: Decay time $\tau$ of the optical mode $b_1$ versus
lg $ N$ at the recombination epoch $(s = 1)$.

\noindent {\bf{Fig. 3}}: Behaviour of the frequency $b$ and
the growth time $\tau = b^{-1}$ for the
acoustic mode $b_2$ at $s = 1$. Maximum unstable region: $2,34
\cdot 10^6 M_\odot \leq M \leq 5,75 \cdot 10^{12} M_\odot$.

\noindent {\bf{Fig. 4}}: Instability valley: The negative $b$-values
of mode $b_2$ versus lg $ N$ and time $s$. The valley
ends at $s = 0,18$ and $N = 2,3 \cdot 10^6$.

\noindent {\bf{Fig. 5}}: Dispersion relation condition (\ref{5.20}) of the
modes $b_1$ and $b_2$ for the typical masses $N = 3,16 \cdot 10^7$
(dotted line) and $N = 10^9$ (solid line) in dependence of $s$.
The condition is fulfilled if the numerical value of $dc$ is
smaller than one, that means in the case of the mode $b_2$ for the
unstable region.

\noindent {\bf{Fig. 6}}: The mode $b_2$ and the associated values of
the amplitude ratio $r$ for
$s = 0,5$ in dependence of $\lg N$.

\noindent {\bf{Fig. 7}}: The behaviour of the matter fluctuation
amplitudes $f(s)$  normalized to its  initial value $f_i$ as
function of $s$ for $N = 4 \cdot 10^6  $ (dotted line) and $N =
7,9 \cdot 10^9$ (solid line), which represent the boundary masses
of the instability valley at the starting point $s = 0,5$ (see
Fig. 6),  in the case of several cosmological models: a) $\Lambda
= 1 \cdot 10^{-56} \,{\rm cm}^{-2}$ , $\epsilon = 0$; b) $\Lambda
= 0$, $\epsilon = 0$; c) $\Lambda = 3,279 \cdot 10^{-56}\, {\rm
cm}^{-2}$,
   $\epsilon = +1$,
d) $\Lambda = 2,206 \cdot 10^{-56}\, {\rm cm}^{-2}$,
   $\epsilon = +1. $

\noindent {\bf{Fig. 8}}: The behaviour of radiation fluctuation
amplitudes $g(s)$ normalized to its initial
value $g_i$ as function of $s$ for the same cases as in Fig. 7.

\noindent {\bf{Fig. 9}}: The behaviour of the matter fluctuation amplitudes
$f(s)$ normalized to its initial value $f_i$ in the case of the
existence of several amounts of nonbaryonic cold dark matter; the
numbers denote the ratio $\delta$ of the nonbaryonic and the
baryonic matter density. Starting point $s = 0,5$ and mass $N =
1,8 \cdot 10^8$.

\noindent {\bf{Fig. 10}}: The behaviour of the radiation fluctuation
amplitudes $g(s)$ normalized to its initial value $g_i$ for the
same cases as in Fig. 9.

\newpage

\alphfig

\begin{figure}[h]
\centering
\psfrag{b}{}
\psfrag{lgN}{}
\includegraphics*[height=5.5cm,width=9cm]{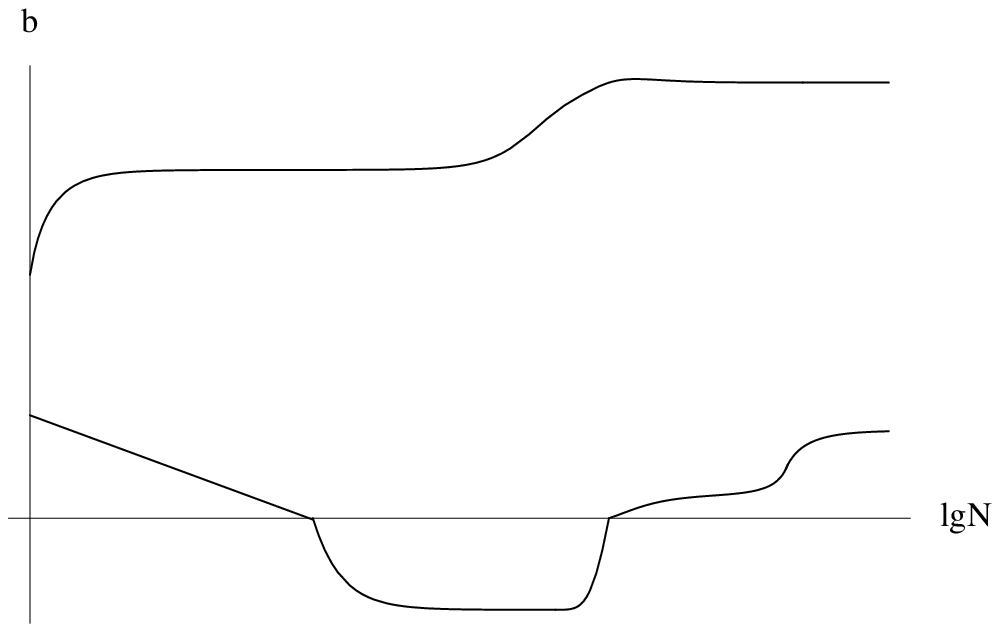}
\unitlength1cm
\multiput(-7.5,0.88)(0,0.2){4}{\line(0,1){0.12}}
\put(-7.68,0.6){\scriptsize $5.3$}
\put(-6.22,0.88){\line(0,1){0.14}}
\put(-6.22,1.1){\scriptsize $6.37$}
\put(-3.55,0.88){\line(0,1){0.14}}
\put(-4.21,1.1){\scriptsize $12.76$}
\put(-4.15,4.81){\line(1,0){0.5}}
\put(-5.78,4.78){{\tiny $\overset{\ess \wedge}{=}$}{\scriptsize $ \, 
1.8 \cdot 10^3 \, \rm{y}$}}
\multiput(-6.15,0.1)(0.2,0){9}{\line(1,0){0.1}}
\put(-7.8,0){{\tiny $\overset{\ess \wedge}{=}$}{\scriptsize $ \, 
1.0 \cdot 10^9 \, \rm{y}$}}
\put(-8.05,1.3){\footnotesize $-$}
\put(-7.05,1.3){\footnotesize $+$}
\put(-1.3,4.52){\footnotesize $-$}
\put(-1.3,1.4){\footnotesize $+$}
\put(-0.8,4.75){$b_1$}
\put(-0.8,1.65){$b_2$}
\put(-0.25,0.85){\small lg\,{\it N}}
\put(-8.8,5.3){$b$}
\put(-0.85,0.935){\vector(1,0){0.15}}
\put(-8.772,4.9){\vector(0,1){0.15}}
       \caption
       { 
}
\end{figure}

\vspace{0.2cm}

\begin{figure}[h]
\centering
\psfrag{r}{}
\psfrag{lgN}{}
\includegraphics*[height=5.5cm,width=9cm]{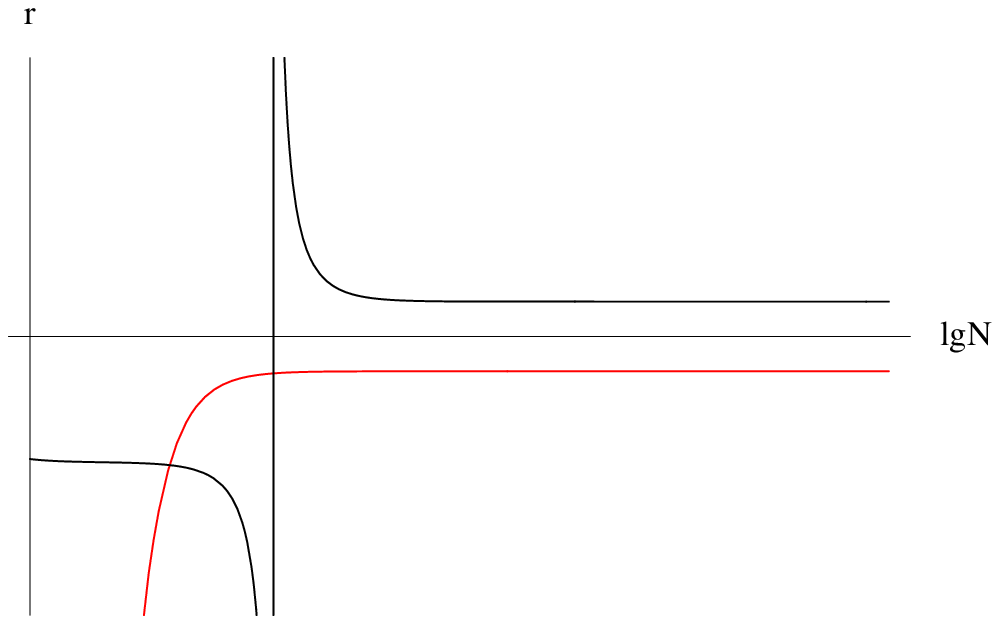}
\unitlength1cm
\put(-7.05,2.6){\scriptsize $5.3$}
\put(-7.35,2.1){\line(1,0){0.35}}
\put(-8.7,2.05){{\scriptsize $- 4.2 \cdot 10^3$}}
\put(-6.32,3.6){\line(1,0){0.2}}
\put(-5.96,3.55){{\scriptsize $4.7 \cdot 10^8$}}
\put(-1.13,1.9){\scriptsize $- 3.09$}
\put(-0.9,2.8){\scriptsize $3.08$}
\put(-4.,1.8){$b_1$}
\put(-4.,2.98){$b_2$}
\put(0.,2.43){\small lg\,{\it N}}
\put(-8.8,5.3){$r$}
\put(-1,2.48){\vector(1,0){0.15}}
\put(-8.764,4.9){\vector(0,1){0.15}}
       \caption
       {
}
\end{figure}

\resetfig

\newpage

\begin{figure}[h]
\centering
\psfrag{t}{\footnotesize $\nms\nms \tau \; [10^{3}\,{\rm y}]$}
\psfrag{lgN}{\footnotesize \;lg\,{\it N}}
           \includegraphics*[height=6cm,width=8cm]{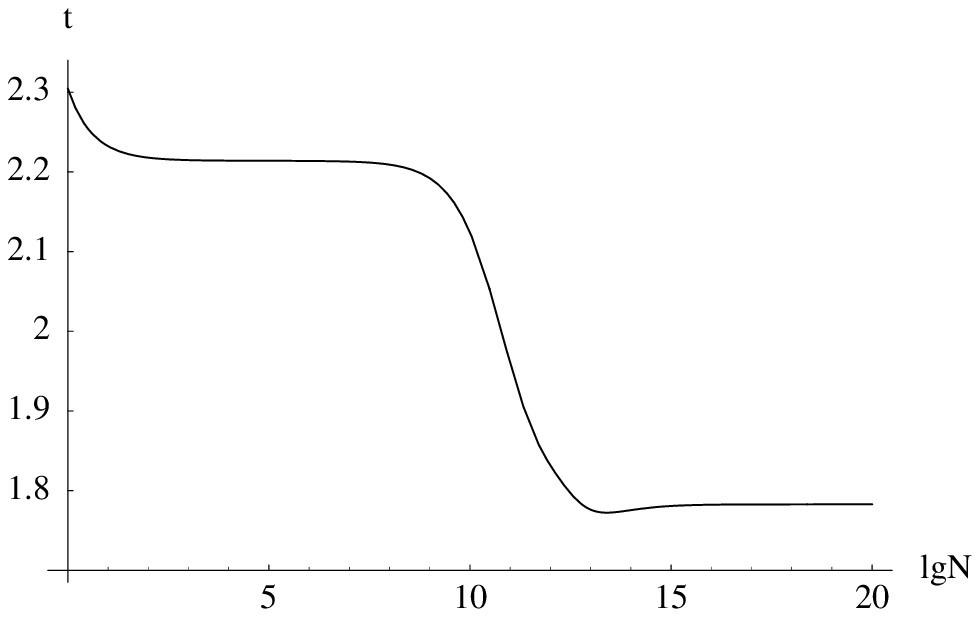}
\unitlength1cm
\put(-0.8,0.475){\vector(1,0){0.15}}
\put(-7.37,5.34){\vector(0,1){0.15}}
           \caption{           
       }
\end{figure}

\vspace{2cm}

\begin{figure}[h]
\centering
\psfrag{tb}{\footnotesize \hspace{-1.25cm} $\nms\nms \tau \; [10^{9}\,{\rm y}]$,
            $b \; [0.5 \cdot 10^{-17} /{\rm sec}]$   }
\psfrag{lgN}{\footnotesize \;lg\,{\it N}}
           \includegraphics*[height=6cm,width=8cm]{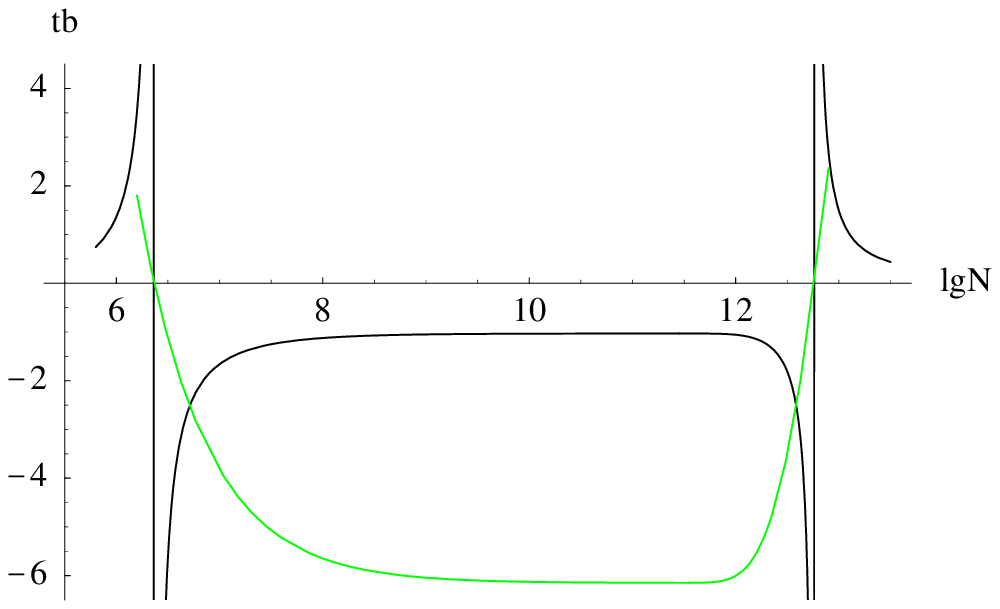}
\unitlength1cm
\put(-4,2.35){$\tau$}
\put(-5.8,1){$b$}
\put(-0.73,3.174){\vector(1,0){0.15}}
\put(-7.475,5.25){\vector(0,1){0.15}}
           \caption{           
       }
\end{figure}

\newpage

\begin{figure}[h]
\centering
\psfrag{6}{}
           \includegraphics*[height=6cm,width=8cm]{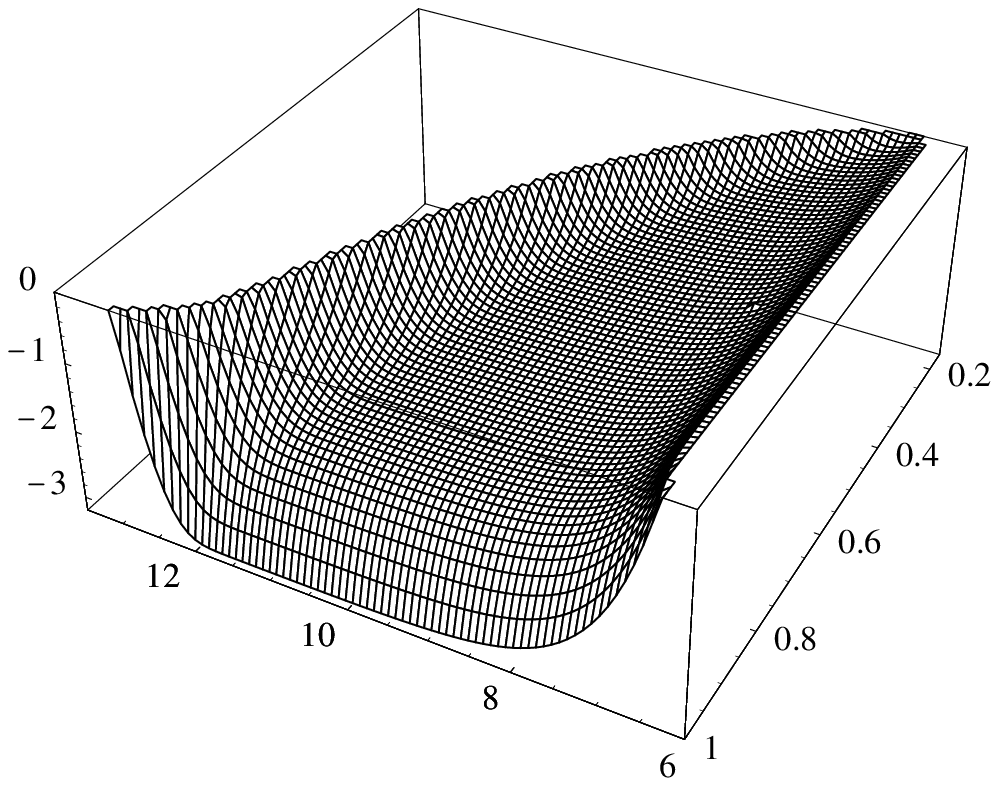}
\unitlength1cm
\put(-0.4,1.9){\footnotesize\it s}
\put(-5.2,0.55){\footnotesize $\nms$ lg\,{\it N}}
\put(-8.1,3){\footnotesize \hspace{-2.1cm} $b \; [10^{-17}/{\rm sec}]$}
           \caption{           
       }
\end{figure}

\begin{figure}[h]
\centering
\psfrag{dc}{\footnotesize dc}
\psfrag{s}{\footnotesize\it s}
           \includegraphics*[height=6cm,width=8cm]{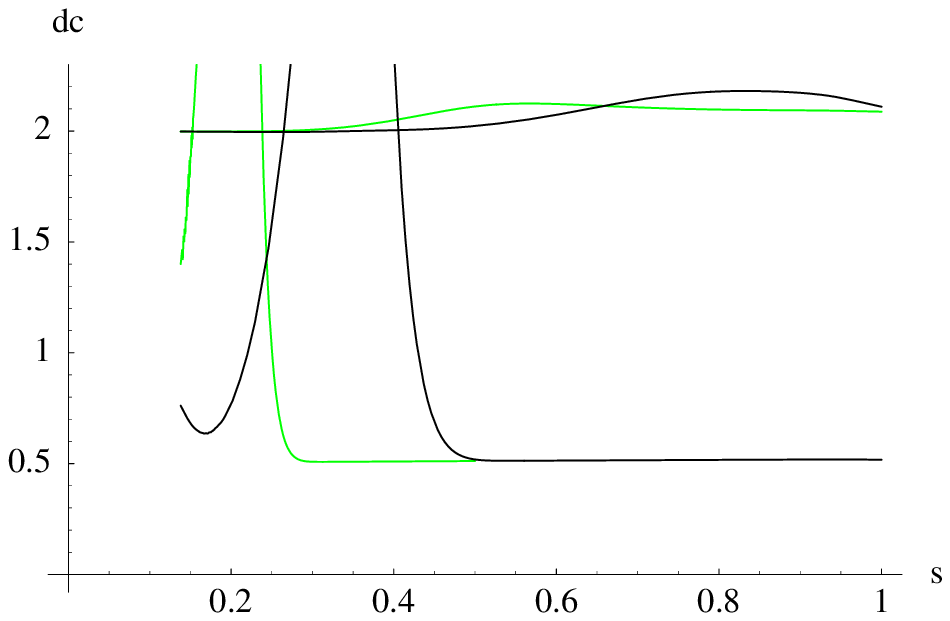}
\unitlength1cm
\put(-1.3,4.4){$b_1$}
\put(-1.3,1.9){$b_2$}
\put(-0.58,0.475){\vector(1,0){0.15}}
\put(-7.255,5.4){\vector(0,1){0.15}}
           \caption{           
       }
\end{figure}

\begin{figure}[h]
\centering
\psfrag{r}{}
\psfrag{m}{\small lg\,{\it N}}
\includegraphics*[height=6.5cm,width=8cm]{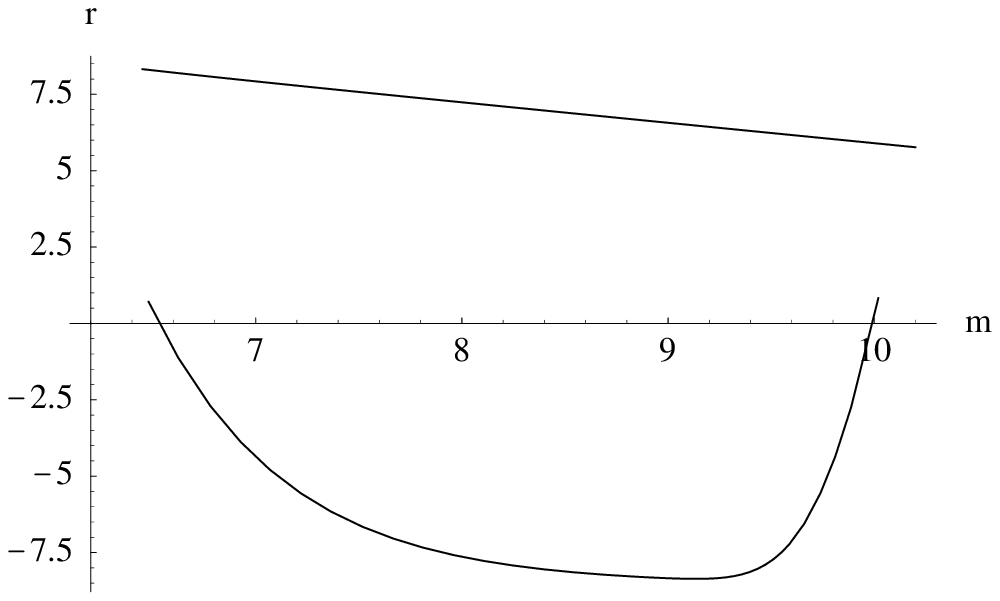}
\unitlength1cm
\put(-3,5.35){$r$}
\put(-1.35,0.5){$b_2$}
\put(-7.6,6.1){ $\lg\,r$}
\put(-7.6,-0.5){$b \; [10^{-18} / {\rm sec}]$}
\put(-0.5,2.943){\vector(1,0){0.15}}
\put(-7.258,5.7){\vector(0,1){0.15}}
       \caption
       {
}
\end{figure}

\begin{figure}[h]
\centering
\psfrag{lgs}{\small lg\,{\it s}}
\psfrag{lgf}{\small lg {\it f/f}$_i$}
\psfrag{1}{\footnotesize  1}
\psfrag{2}{\footnotesize  2}
\psfrag{3}{\footnotesize  3}
\psfrag{4}{\footnotesize  4}
\psfrag{0}{\footnotesize 0}
\psfrag{0.5}{\footnotesize 0.5}
\psfrag{1.5}{\footnotesize 1.5}
\psfrag{2.5}{\footnotesize 2.5}
\includegraphics*[height=7.5cm,width=10.5cm]{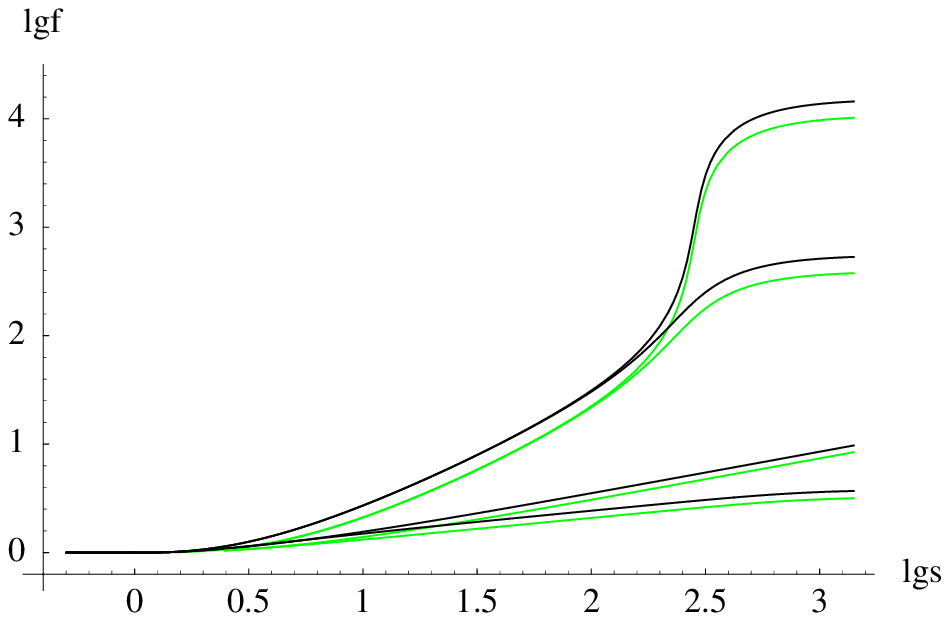}
\unitlength1cm
\put(-1,6.25){\small d)} 
\put(-1,4.35){\small c)} 
\put(-1,2.1){\small b)} 
\put(-1,1.4){\small a)} 
\put(-1.06,0.6){\vector(1,0){0.15}}
\put(-9.78,6.78){\vector(0,1){0.15}}
\caption
{}
\end{figure}

\begin{figure}[h]
\centering
\psfrag{lgs}{\small lg\,{\it s}}
\psfrag{lgg}{\small lg {\it g/g}$_i$}
\psfrag{1}{\footnotesize  1}
\psfrag{2}{\footnotesize  2}
\psfrag{3}{\footnotesize  3}
\psfrag{-}{\footnotesize  -}
\psfrag{0}{\footnotesize 0}
\psfrag{0.5}{\footnotesize 0.5}
\psfrag{1.5}{\footnotesize 1.5}
\psfrag{2.5}{\footnotesize 2.5}
\includegraphics*[height=7.5cm,width=10.5cm]{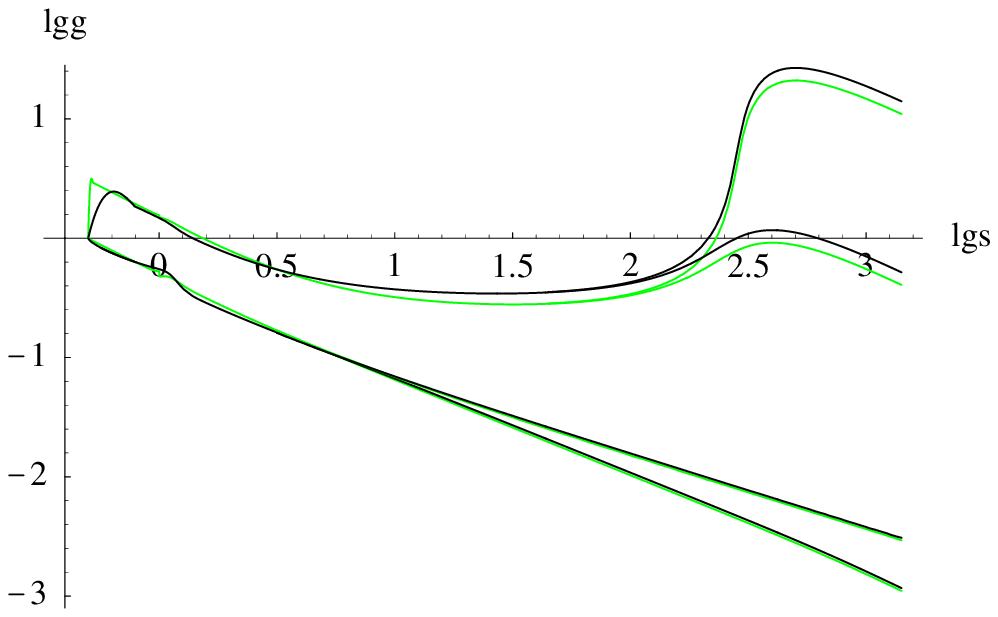}
\unitlength1cm
\put(-0.7,6.2){\small d)} 
\put(-0.7,3.9){\small c)} 
\put(-0.7,0.9){\small b)} 
\put(-0.7,0.1){\small a)} 
\put(-0.85,4.62){\vector(1,0){0.15}}
\put(-9.812,6.72){\vector(0,1){0.15}}
\caption
{
}
\end{figure}

\begin{figure}[h]
\centering
\psfrag{lgs}{\small   lg\,{\it s}}
\psfrag{lfg}{\small lg {\it f/f}$_i$}
\psfrag{1}{\footnotesize  1}
\psfrag{2}{\footnotesize  2}
\psfrag{3}{\footnotesize  3}
\psfrag{4}{\footnotesize  4}
\psfrag{0}{\footnotesize 0}
\psfrag{0.5}{\footnotesize 0.5}
\psfrag{1.5}{\footnotesize 1.5}
\psfrag{2.5}{\footnotesize 2.5}
\includegraphics*[height=7.5cm,width=10.5cm]{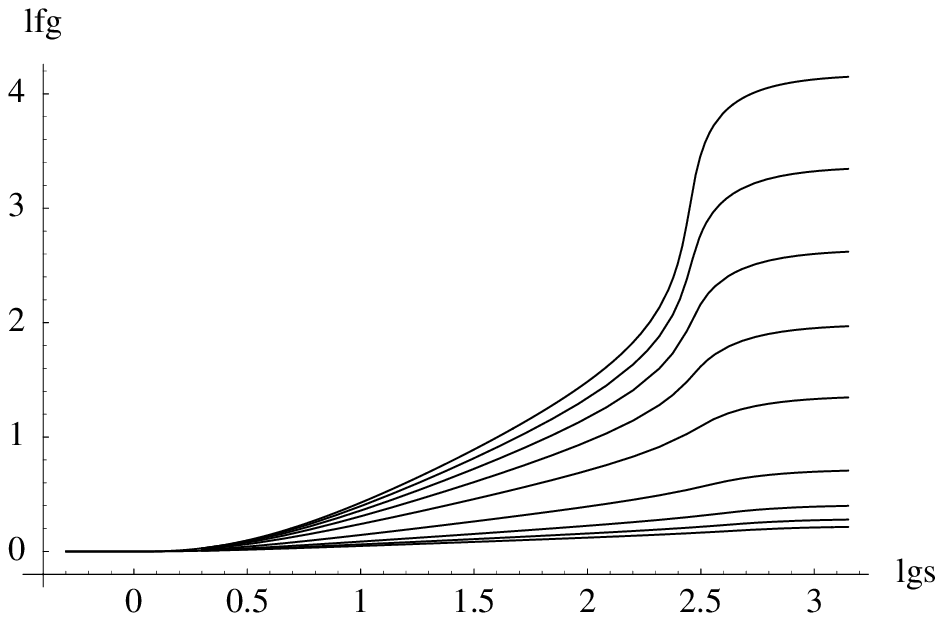}
\unitlength1cm
\put(-1,6.53){\footnotesize 0} 
\put(-1,5.42){\footnotesize 0.1} 
\put(-1,4.39){\footnotesize 0.25} 
\put(-1,3.51){\footnotesize 0.5} 
\put(-1,2.65){\footnotesize 1} 
\put(-1,1.78){\footnotesize 2.5} 
\put(-0.4,1.35){\scriptsize 5} 
\put(-1,1.15){\footnotesize 7.5} 
\put(-0.4,1.07){\scriptsize 10} 
\put(-1.1,0.59){\vector(1,0){0.15}}
\put(-9.763,6.78){\vector(0,1){0.15}}
\caption
{}
\end{figure}

\begin{figure}[h]
\centering
\psfrag{lgs}{\small lg\,{\it s}}
\psfrag{lgg}{\small lg {\it g/g}$_i$}
\psfrag{1}{\footnotesize  1}
\psfrag{2}{\footnotesize  2}
\psfrag{3}{}
\psfrag{-}{\footnotesize  -}
\psfrag{0}{\footnotesize 0}
\psfrag{0.5}{}
\psfrag{1.5}{\footnotesize 1.5}
\psfrag{2.5}{}
\includegraphics*[height=7.5cm,width=10.5cm]{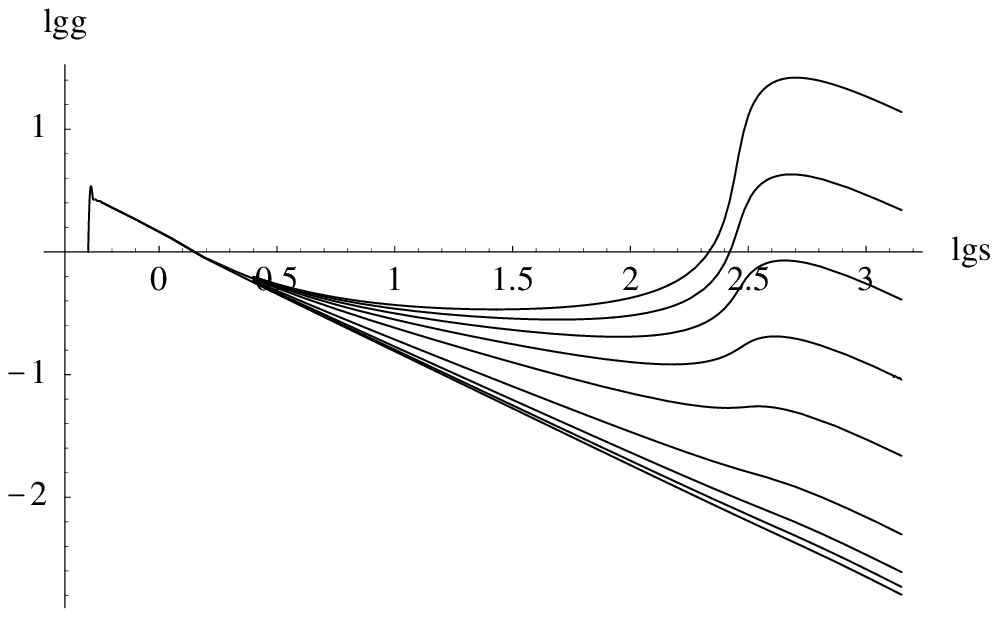}
\unitlength1cm
\put(-7.8,4.65){\footnotesize 0.5}
\put(-2.64,4.65){\footnotesize 2.5}
\put(-1.47,4.65){\footnotesize 3}
\put(-0.8,5.93){\footnotesize 0}
\put(-0.8,4.8){\footnotesize 0.1}  
\put(-0.8,3.68){\footnotesize 0.25} 
\put(-0.8,2.7){\footnotesize 0.5} 
\put(-0.8,1.79){\footnotesize 1} 
\put(-0.8,0.82){\footnotesize 2.5} 
\put(-0.25,0.33){\scriptsize 5} 
\put(-0.8,0.22){\footnotesize 7.5} 
\put(-0.25,0.05){\scriptsize 10} 
\caption
{
}
\end{figure}

\end{document}